\documentclass[12pt]{JHEP3}

\usepackage[english]{babel}
\usepackage[T1]{fontenc}
\usepackage[latin1]{inputenc}
\usepackage{cite}

\usepackage{amsmath}
\usepackage{bm}
\usepackage{mathrsfs}
\usepackage{tabularx}
\usepackage{amssymb}
\usepackage{paralist}
\usepackage[active]{srcltx}
\usepackage[mathscr]{eucal}
\usepackage{caption}

\newcommand{\bra}[1]{\left<\,{#1}\,\right|}
\newcommand{\ket}[1]{\left|\,{#1}\,\right>}
\newcommand{\bk}[2]{\left<\,{#1} \mid {#2}\,\right>}

\preprint{arXiv:1007.4384 [hep-th]}

\title{Higher spin interactions with scalar matter\\ on constant curvature spacetimes: \\ \large{Conserved current and cubic coupling generating functions}}

\author{Xavier Bekaert ~~and~~ Elisa Meunier\\
Laboratoire de Math\'ematiques et Physique Th\'eorique\\
Unit\'e Mixte de Recherche $6083$ du CNRS\\
F\'ed\'eration de Recherche $2964$ Denis Poisson\\
Universit\'e Fran\c{c}ois Rabelais, Parc de Grandmount\\
37200 Tours, France\\

\email{bekaert@lmpt.univ-tours.fr},\email{ elisa.meunier@lmpt.univ-tours.fr}}

\abstract{Cubic couplings between a complex scalar field
and a tower of symmetric tensor gauge fields of all ranks are investigated on any constant curvature spacetime of dimension $d\geqslant 3$.
Following Noether's method, the gauge fields interact with the scalar field via minimal coupling to the conserved currents.
A symmetric conserved current, bilinear in the scalar field and containing up to $r$ derivatives, is obtained for any rank $r\geqslant 1$ from its flat spacetime counterpart in dimension $d+1$, via a radial dimensional reduction valid precisely for the mass-square domain of unitarity in (anti) de Sitter spacetime of dimension $d$.
The infinite collection of conserved currents and cubic vertices are summarized in a compact form
by making use of generating functions and of the Weyl/Wigner quantization on constant curvature spaces.}


\begin{document}


\section{Introduction}

Principal bundles and Riemannian manifolds provide the right geometrical frameworks for describing the interactions between gauge fields with respective spin one and two. However, despite remarkable results on the interactions between higher spin gauge fields their underlying geometrical and physical first principles remain elusive. Although a higher-spin generalization of gravity is available through the frame-like formulation of Vasiliev (see \textit{e.g.} \cite{Vasiliev:2004qz} for some reviews) extending the Cartan-Weyl formulation of general relativity, the first principles analogous to the parallel transport and to the local affine covariance on the geometrical side, or to the gauge and equivalence principles on the physical side, still remain mysterious.
The latter physical principles, underlying the low-spin interactions, are best displayed in the minimal couplings between matter and gauge fields, so higher-spin generalizations thereof might be a proper place to look for inspiration. Specifically, one will concentrate here on a toy model where matter is represented by a complex scalar field. This simplest example already proved to highlight most of the key features of the more intricate general couplings between fields of non-vanishing spins.

The Noether (\textit{i.e.} minimal) cubic couplings between a complex scalar matter field and a collection of higher-spin tensor gauge fields have already been investigated in the metric-like formulation on Minkowski \cite{Berends:1985xx,Calimanesti,Bekaert:2009ud,Fotopoulos} and anti de Sitter \cite{Fotopoulos,Ruhl,Fotopoulos:2010nj} spacetimes (see also the recent work \cite{Zinoviev:2010cr}
in the frame-like formulation).
The Noether cubic interaction between a complex scalar field and a tensor gauge field takes place through a symmetric current, quadratic in the scalar field and conserved at linearized level. 
By construction, such models are consistent from quadratic order in the gauge and matter
fields up to cubic couplings of two scalar and the gauge fields.
The present paper may be thought as a first step towards a complete generalization to any constant curvature spacetime of the results obtained in \cite{Bekaert:2009ud} on Minkowski spacetime. Our strategy is to derive the non-zero curvature formulas from the flat spacetime results by performing a so-called ``radial dimensional reduction'' \cite{Biswas:2002nk} also called ``ambient space formulation'', \textit{i.e.} by making use of the usual isometric embedding of (anti) de Sitter spacetime as a codimension one hyperboloid inside a flat auxilliary space. The basic idea goes back to an early work of Dirac \cite{Dirac:1935zz}. In the late seventies, the ambient formulation had already been used by Fronsdal \cite{Fronsdal:1978vb} in the context of higher-spin gauge theories and, by now, this technique has become standard and has found a large number of applications in this area (see \textit{e.g.} \cite{Metsaev,Bekaert:2003uc,Fotopoulos:2007yq,Francia:2008hd}). 

The plan of the paper is as follows: In order to be self contained, the framework {presented} in \cite{Bekaert:2009ud} (\textit{i.e.} the various generating functions relevant for the Noether method in the case of gauge/matter couplings) is reviewed in Section \ref{Noethermethod}, but from a slightly more general viewpoint (allowing for curved background) suited to the present analysis. In the section \ref{versus}, a dictionary between two formulations (the intrinsic and the ambient ones) of fields on non-zero constant-curvature spacetimes is provided. The treatment is uniform with respect to the signature and to the sign of the scalar curvature, in order to incorporate both (anti) de Sitter spacetimes and their Euclidean counterpart, \textit{i.e.} hyperspheres (hyperbolic spaces). The infinite set of conserved currents bilinear in a free complex scalar field are presented in Section \ref{conscur}. The corresponding Noether cubic vertex is given in Section \ref{Noethercoupl} and is written in a  compact form
by making use of Weyl/Wigner symbol calculus, which enables the explicit computation of the non-Abelian gauge symmetry deformation. In the last section \ref{conclusion}, our main results are summarized. Some possible extensions thereof are also suggested and motivated.
Eventually, the paper ends with an appendix where a technical proof is presented in details.

\section{Noether method}\label{Noethermethod}

Let $\mathcal M _d$ be a (pseudo) Riemannian manifold of dimension $d$ endowed with a metric $g_{\mu\nu}$ (Minuscule Greek indices $\mu,\nu,\ldots $ will take $d$ values and they will be lowered or raised via the metric or its inverse) and its associated Levi-Civita connection $\nabla_\mu\,$.

A \textit{symmetric conserved current} of rank $r\geqslant 1$ is a real contravariant symmetric tensor field $j^{\,\mu_1\ldots\,\mu_r}(x)$ on $\mathcal M _d$ obeying to the conservation law 
\begin{equation}
\nabla_{\mu_1}j^{\mu_1\ldots\,\mu_r}(x)\approx 0\,.
\label{conservationlaws}
\end{equation}
where the ``weak equality'' symbol $\approx\,$ stands for ``equal on-mass-shell,'' \textit{i.e.} modulo terms proportional to the Euler-Lagrange equations.
A \textit{generating function of conserved currents} is a real function $j(x,p)$ on the phase space $T^*\mathcal{M} _d$ which is (i) a formal power series in the momenta and (ii) such that
\begin{equation}
\left(\nabla_\mu\frac{\partial}{\partial p_\mu}\right)\,j(x,p)\approx 0\,.
\label{conservationlaw}
\end{equation}
This terminology follows from the fact that all the coefficients of order $r\geqslant 1$ in the power expansion of the generating function
\begin{equation}
j(x,p)\,=\,\sum\limits_{r\geqslant 0}\frac{1}{r!}\,j^{\mu_1\ldots\,\mu_r}(x)\,p_{\mu_1}\ldots  p_{\mu_r}
\label{j}
\end{equation}
are all symmetric conserved currents by means of (\ref{conservationlaw}).

A \textit{symmetric tensor gauge field} of rank $r\geqslant 1$ is a real covariant symmetric tensor field $h_{\mu_1\ldots \mu_r}(x)$ on $\mathcal M _d$ whose gauge transformations are of the form \cite{Fronsdal:1978vb}
\begin{equation}
\delta_\varepsilon h_{\mu_1\ldots \mu_r}(x)\,=\,r\,\nabla_{(\mu_1}\varepsilon_{\mu_2\ldots \mu_r)}(x)\,+\,{\cal O}(h)\,,
\label{gaugetransfo}
\end{equation}
where the gauge parameter $\varepsilon_{\mu_1\ldots \mu_{r-1}}(x)$ is a covariant symmetric tensor field of rank $r-1$,
the round bracket denotes complete symmetrization with weight one, \textit{i.e.} $h_{(\mu_1\ldots \mu_r)}=h_{\mu_1\ldots \mu_r}$ (remark: the tensor is symmetric by hypothesis) and ${\cal O}(h)$ stands for terms of order one or more in the gauge fields.
For lower ranks $r=1$ or $2\,$, the transformation (\ref{gaugetransfo}) either corresponds to the $\mathop{\rm {}U}(1)$ gauge transformation of the vector ($r=1$) gauge field or to the linearized diffeomorphisms of the metric ($r=2$). By comparison with the spin-two case, this formulation of higher-spin gauge fields is {often} called ``metric-like'' (in order to draw the distinction with the ``frame-like'' version where the gauge field is not completely symmetric).
A \textit{generating function of gauge fields} is a real function $h(x,v)$ on the configuration space $T\mathcal{M} _d$ (i) which is a formal power series in the velocities and (ii) whose gauge transformations are
\begin{equation}
\delta_\varepsilon h(x,v)\,=\,\left(v^\mu \nabla_\mu\right)\,\varepsilon(x,v)\,+\,{\cal O}(h)\,,
\label{Fronsdalgtransfo}
\end{equation}
where $\varepsilon(x,v)$ is also a formal power series in the velocities.
The nomenclature follows from the fact that all the coefficients of order $r\geqslant 1$ in the power expansion of the generating function
\begin{equation}
h(x,v)\,=\,\sum\limits_{r\geqslant 0}\frac{1}{r!}\,h_{\mu_1\ldots\, \mu_r}(x)\,v^{\mu_1}\ldots  v^{\mu_r}
\label{h}
\end{equation}
are all symmetric tensor gauge fields due to (\ref{Fronsdalgtransfo}) with 
\begin{equation}
\varepsilon(x,v)\,=\,\sum\limits_{t\geqslant 0}\frac{1}{t!}\,\varepsilon_{\mu_1\ldots\, \mu_t}(x)\,v^{\mu_1}\ldots  v^{\mu_t}\,.
\end{equation}

In the context of Noether couplings, the ``velocities'' $v^\mu$ and ``momenta'' $p_\nu$ are interpreted as mere auxiliary variables and can be assumed to be dimensionless. 
Let us introduce a non-degenerate bilinear pairing $\ll \| \gg$ between smooth functions $h(x,v)$ and $j(x,p)$ on the configuration and phase spaces respectively,
\begin{equation}
\ll h\,\|\,j \gg\,\,:=\,\int_{\mathcal M _d} d^dx\,\sqrt{|\,g|} \,\left.\exp\left(\,\frac{{\partial}}{\partial v^{\mu}}\,\frac{{\partial}}{\partial p_{\mu}}\right)h(x,v)\,j(x,p)\,\right|_{v=p=0}\,\,.
\label{pairing}
\end{equation}
If $j$ and $h$ are (formal) power series of the form (\ref{j}) and (\ref{h}) then the pairing (\ref{pairing}) can be interpreted as the series
\begin{equation}
\ll h\,\|\,j \gg\,\,=\,\sum\limits_{r\geqslant 0}\,\frac{1}{r!}\,\int_{\mathcal M _d} d^dx\,\sqrt{|\,g|}\,\,h_{\mu_1\ldots \mu_r}(x)\,j^{\,\mu_1\ldots \mu_r}(x)\,.
\label{pairingsum}
\end{equation}

Let us denote by $\ddagger$ the adjoint operation for the pairing (\ref{pairing}) in the sense that 
\begin{equation}
\ll \Hat{\Hat{O}}\,h\,\|\,j \gg\,=\,\ll h\,\|\,\Hat{\Hat{O}}^\ddagger\, j \gg\,,
\end{equation}
where $\Hat{\Hat{O}}$ is an operator acting on the vector space of functions on configuration space (the double hat stands for ``second quantization'' in the sense that the operator acts on symbols of ``first quantized'' observables).
Notice that $(v^\mu)^\ddagger={\partial}/{\partial p_\mu}$ and $\nabla_\mu^\ddagger=-\nabla_\mu$ imply the useful relation
\begin{equation}
\left(v^\mu\, \nabla_\mu\right)^\ddagger\,=\,-\,\left(\nabla_\mu\,\frac{\partial}{\partial p_\mu}\right).
\label{adjoint}
\end{equation}

The \textit{matter action} is a functional $S_0[\phi]$ of some matter fields collectively denoted by $\phi\,$. The Euler-Lagrange equations of these matter fields is such that there exists some conserved current $j^{\mu_1\ldots \mu_r}[\,\phi(x)\,]\,$.   
The Noether method for introducing interactions is essentially the ``minimal'' coupling between a gauge field $h_{\mu_1\ldots \mu_r}(x)$ and a conserved current $j^{\mu_1\ldots \mu_r}[\,\phi(x)\,]$ of the same rank.
Accordingly, the \textit{Noether interaction} between gauge fields and conserved currents is the functional defined as the pairing between their generating functions
\begin{equation}
S_1[\phi,h]\,:=\,\,\ll h\,\|\, j\gg\,\,=\, \sum\limits_{r\geqslant 0}\,\frac{1}{r!}\,\int_{\mathcal M _d} d^dx\,\sqrt{|\,g|}\,\,h_{\mu_1\ldots \mu_r}(x)\,j^{\mu_1\ldots \mu_r}(x)\,,
\label{Noetherinteraction}
\end{equation}
where (\ref{pairingsum}) has been used.
Let us assume that there exists a gauge invariant action $S[\phi,h]$ whose power expansion in the gauge fields starts as follows
\begin{equation}
S\,[\phi,h]\,=\,S_0[\phi]\,+\,S_1[\phi,h]\,+\,S_2[\phi,h]\,+\,{\cal O}(h^3)\,.
\label{actionexpansion}
\end{equation}
The gauge variation of the Noether interaction (\ref{Noetherinteraction}) under (\ref{Fronsdalgtransfo}),
\begin{equation}
\delta_\varepsilon S_1[\phi,h]\,=\,\,\ll \delta_\varepsilon h\,\|\, j\gg\,+\,{\cal O}(h)\,,
\end{equation}
is at least of order one in the gauge fields when the equations of motion for the matter sector are obeyed, 
\begin{equation}
\delta_\varepsilon S_1[\phi,h]\,\approx\,{\cal O}(h)\,,
\label{gtransfoaction}
\end{equation}
because the properties (\ref{conservationlaw}) and (\ref{adjoint}) imply that
\begin{equation}
\ll \Big(v^\mu \nabla_\mu\Big)\,\varepsilon\,\|\, j\gg\,=\,-\,\ll \varepsilon\,\|\, \Big(\nabla_\mu\frac{\partial}{\partial p_\mu}\Big)\,j\gg\,\,\approx\,0\,.
\label{gauge/conservation}
\end{equation}
Actually, the crucial property (\ref{gtransfoaction}) works term by term since
\begin{eqnarray}
&&\int_{\mathcal M _d} d^dx\,\sqrt{|\,g|}\,\,\nabla_{\mu_1}\varepsilon_{\mu_2\ldots \mu_r}(x)\,j^{\mu_1\ldots \mu_r}(x)\nonumber\\
&&\quad=\,-\int_{\mathcal M _d} d^dx\,\sqrt{|\,g|}\,\,\varepsilon_{\mu_2\ldots \mu_r}(x)\,\nabla_{\mu_1}j^{\mu_1\ldots \mu_r}(x)\,\approx\,0\,.
\end{eqnarray}
The equation (\ref{gtransfoaction}) implies that the action (\ref{actionexpansion}) might indeed be gauge-invariant at lowest order in the gauge fields because the terms in $\delta_\varepsilon S_1[\phi,h]$ that are proportional to the Euler-Lagrange equations $\delta S_0/\delta\phi$ of the matter sector could be compensated by the {variation $\delta_\varepsilon S_0[\phi]$ of the matter action under} a gauge transformation $\delta_\varepsilon\phi$ of the matter fields, independent of the gauge fields $h$ and linear in the matter fields $\phi\,$, such that
\begin{equation}
\delta_\varepsilon \Big(\,S_0[\phi]+ S_1[\phi,h]\,\Big)\,=\,{\cal O}(h)\,.
\label{lowestorder}
\end{equation}
This possibility will be assumed from now on.

\vspace{1mm}
A \textit{Killing tensor field} of rank $r-1\geqslant 0$ on $\mathcal{M} _d$ is a real covariant symmetric tensor field $\overline{\varepsilon}_{\mu_1\ldots \mu_{r-1}}(x)$ 
solution of the generalized Killing equation
\begin{equation}
\nabla_{(\mu_1}\overline{\varepsilon}_{\mu_2\ldots \mu_r)}(x)=0\,.
\end{equation}
A \textit{generating function of Killing fields} is a function $\overline{\varepsilon}(x,v)$ on the configuration space $T\mathcal{M} _d$ which is (i) a formal power series in the velocities and (ii) such that $(v^\mu \nabla_\mu)\overline{\varepsilon}(x,v)=0\,$.
Then the coefficients in the power series
\begin{equation}
\overline{\varepsilon}(x,v)\,=\,\sum\limits_{t\geqslant 0}\frac{1}{t!}\,\overline{\varepsilon}_{\mu_1\ldots \mu_t}(x)\,v^{\mu_1}\ldots  v^{\mu_t}
\end{equation}
are all Killing tensor fields on $\mathcal{M} _d\,$.
The variation (\ref{gaugetransfo}) of the gauge field vanishes at lowest order if the gauge parameter is a Killing tensor field. Therefore the corresponding transformation $\delta_{\overline\varepsilon}\phi$ of the matter fields is a rigid symmetry of the matter action $S_0[\phi]\,$:
\begin{equation}\delta_{\overline\varepsilon}S_0[\phi]=-\,\delta_{\overline\varepsilon}S_1[\phi,h]\,\big|_{_{h=0}}=0\,,\end{equation}
due to (\ref{lowestorder}) and the fact that $\delta_\varepsilon\phi$ is independent of the gauge fields.
In turn, this shows that the conserved current $j^{\mu_1\ldots \mu_r}[\,\phi(x)\,]$ must be equal, on-shell and modulo a trivial conserved current (sometimes called an ``improvement''), to the Noether current associated with the latter rigid symmetry $\delta_{\overline\varepsilon}\phi$ of the matter action $S_0[\phi]\,$.
Killing tensor fields on constant curvature spacetimes and their link with higher-spin gauge theories were discussed in more details in \cite{Bekaert:2005ka} and references therein.

\section{Ambient \textit{versus} intrinsic formulations}\label{versus}

\subsection{Constant curvature manifolds}

Let ${\mathbb R}^{D}$ be the flat space of dimension $D\geqslant 4$ parametrized by Cartesian coordinates $X^A$ (Capital Latin indices $A,B,\ldots $ will span $D$ values) and endowed with a non-degenerate diagonal metric $\eta_{AB}$ that will be used to raise and lower Capital Latin indices.
It will be called the \textit{ambient} space. The inner product will be denoted as $X\cdot Y\,:=\,\eta_{AB}\,X^A\,Y^B$ (and $X^2\,:=\,\eta_{AB}\,X^A\,X^B$).
Let $\mathcal{M} _d$ be the non-degenerate quadric of dimension $d:=D-1$ defined by the equation 
$X^2\,=\,\pm\,R^2\,,$ where $R\ne0$ is its curvature radius. The sign is fixed in the previous expression, but the $\pm$ has been included to deal with both cases at once. From now on, the $\pm$ and $\mp$ symbols in the subsequent formulae will always correspond to this respective choice of sign. For instance, the (pseudo) Riemannian manifold $\mathcal{M}_d$ has constant scalar curvature equal to ${\cal R}= \pm\,d(d-1)/R^2$.

Let us denote by $x^\mu$ a set of coordinates on $\mathcal{M}_d$ with length dimension (in the sense that they scale in the same way as the Cartesian coordinates $X^A$). They will be called \textit{intrinsic} coordinates.
One considers an isometric smooth embedding
\begin{eqnarray}
i \,:\,\mathcal{M}_d\,\hookrightarrow\,\mathbb{R}_0^D\, : \, x^\mu \,\longmapsto X^A(x^\mu)
\label{inclusion}
\end{eqnarray}
of the codimension-one quadric $\mathcal{M} _d$ inside the open submanifold $\mathbb{R}_0^D\subset {\mathbb R}^{D}$ defined by
\begin{equation}\mathbb{R}_0^D\,\,:=\,\,\{\,X^A\in {\mathbb R}^{D}\,:\,\pm X^2>0\,\}\,.\end{equation}
The (pseudo) ``spherical'' coordinates $(\rho,y^\mu)$ collect the ``radial'' coordinate $\rho:=\sqrt{\pm X^2}$ together with the
dimensionless ``angular'' coordinates $y^\mu$($:=x^\mu/R$) of the radial projection of the given point of $\mathbb{R}_0^D$ on $X^2=\pm 1$. This coordinate system covers the manifold ${\mathbb R}_0^{D}$. The submanifold $\mathcal{M}_d\subset \mathbb{R}_0^D$ is simply the locus such that $\rho=R$.

\subsection{Tensor fields}

Let $\mathscr{X}_r(\mathcal{M} _d)$ denote the space of smooth rank-$r$ covariant tensor fields $t_{\mu_1\ldots \mu_r}(x)$ on $\mathcal{M} _d$
and $\mathscr{X}_r(\mathbb{R}_0^D)$ the space of smooth rank-$r$ covariant tensor fields $T_{A_1\ldots A_r}(X)$ on $\mathbb{R}_0^D$, both with values in $\mathbb R$ (or $\mathbb C$ in general).
The pull-back
\begin{eqnarray}
i^*&:&\mathscr{X}_r(\mathbb{R}_0^D)\,\rightarrow\,\mathscr{X}_r(\mathcal{M} _d)\nonumber\\
&:&
T_{A_1\ldots A_r}(X)\,\longmapsto\,t_{\mu_1\ldots \mu_r}(x)=\dfrac{\partial X^{A_1}(x)}{\partial x^{\mu_1}}\cdots \dfrac{\partial X^{A_r}(x)}{\partial x^{\mu_r}}\,T_{A_1\ldots A_r}\left(X(x)\right) 
\label{tenseur1}
\end{eqnarray}
is surjective but not injective.
However, there exists a nice isomorphism between the space $\mathscr{X}_r(\mathcal{M} _d)$ of rank-$r$ tensor fields on $\mathcal{M} _d$
and the subspace of rank-$r$ tensor on $\mathbb{R}_0^D$ that are:
\begin{itemize}
  \item[(i)] \textit{homogeneous} of fixed non-zero homogeneity degree (say $k\in {\mathbb C}_0$),
\begin{equation}T_{A_1\ldots A_r}(\lambda X)\,=\,\lambda ^k\,T_{A_1\ldots A_r}(X)\,,\qquad \forall \lambda\in {\mathbb C}_0\,.\end{equation}
	\item[(ii)] \textit{tangent} to the constant $\rho$ submanifolds, \textit{i.e.}
\begin{equation}
X^{A_i}\,T_{A_1\ldots  A_i\ldots  A_r}(X) \,=\,0
\label{normal}
\end{equation}
\end{itemize}
This isomorphism was explained in details by Fronsdal in \cite{Fronsdal:1978vb} but one may review the construction as follows: 

The condition (i) is best understood for scalar fields ($r=0$) since the condition (ii) is absent. On the one hand, the restriction to $\mathcal{M} _d$ maps any function $\Phi(X)$ on ${\mathbb R}_0^{D}$ to the function on $\mathcal{M} _d$ given by \footnote{With a slight abuse of notation, we denote by $\Phi(\rho,x^\mu)$ the pull-back  $\Phi\left(X^A(\rho,x^\mu)\right)$. Moreover, in the sequel we will also frequently denote by $\phi(x^\mu)$ the function $\phi\left(y^\mu(x)\right)$.}
\begin{eqnarray}
\phi(y^\mu)\,=\,\Phi(\rho,y^\mu)|_{\rho\,=\,R\,}=\,\Phi(R,\,y^\mu)\,=\,\Phi(X^A)|_{X^2=R^2}\,.
\label{defphi}
\end{eqnarray}
On the other hand, to any function $\phi(x)$ on $\mathcal{M} _d$ one may associate a homogeneous function $\Phi(X)$ of degree $k$ on ${\mathbb R}_0^{D}$ given by  
\begin{equation}
\Phi(X^A)\,=\,\Phi(\rho,\,y^\mu)\,=\, \left(\dfrac{\rho}{R}\right)^k\,\Phi\left(R,\,y^\mu\right)\,=\, \left(\dfrac{\rho}{R}\right)^k\,\phi\left(y^\mu\right),
\label{homh}
\end{equation}
whose restriction on $\mathcal{M} _d$ reproduces $\phi(y)$ as in (\ref{defphi}). This function $\Phi(X)$ is indeed of homogeneity degree $k$ in $X$ (or in $\rho$), 
\begin{eqnarray}
\Phi(\lambda X)\,=\,\lambda^k\,\Phi(X)\,,
\label{homogeneite}
\end{eqnarray}
since $X^{\prime A}=\lambda X^A$ is equivalent to $\rho^\prime=\lambda \rho$ and $y^{\prime\mu}=y^\mu$ (because the dimensionless angular coordinates do not scale with respect to the Cartesian coordinates $X^A$). 
The fancy terminology ``radial dimenional reduction'' \cite{Biswas:2002nk} comes from the analogy of (\ref{homh}) with a usual dimensional reduction ansatz along the direction parametrized by $z:=\log(\rho/R)$ since then $\Phi(X^A)\,=\, \exp(kz)\,\phi(y^\mu)$ looks like a Fourier mode ansatz (when $k$ is pure imaginary). More comments on this point will be made further below.

The condition (ii) takes into account the projection of the components of the ambient tensor $T_{A_1\ldots  A_r}(X)$ on the coordinate basis $\partial/\partial x^\mu$ on each tangent space
through the pull-back formula (\ref{tenseur1}). The standard condition
\begin{eqnarray}
\dfrac{\partial X}{\partial x^\mu}\,\cdot\,X\,=\,0
\label{transverse}
\end{eqnarray}
implies that the kernel of the pull-back (\ref{tenseur1}) for ambient vector fields $V^A(X)$ is spanned by the radial vector fields, \textit{i.e.} such that $V^A(X)= X^A \Phi(X)$.
Therefore, the space of tangent tensors $t_{\mu_1\ldots \mu_r}(x)\in T^*_q\mathcal{M} _d$ at a point $q\in \mathcal{M} _d$ of Cartesian coordinates $X^A$ is isomorphic to the space of ambient tensors
$T_{A_1\ldots A_r}(X)\in T^*_q{\mathbb R}_0^{D}$ that are tangent to $\mathcal{M} _d$ at the same point $q\in \mathcal{M} _d\subset{\mathbb R}_0^{D}$ or, equivalently, that are are normal to the radial direction, \textit{i.e.} they satisfy to (\ref{normal}). 

The operator of orthogonal projection of ambient vectors on the tangent bundle $T\mathcal{M} _d$ is equal to 
\begin{equation}
\mathcal P_A^B\,=\, \delta_A^B \,-\, \dfrac{X_A X^B}{X^2}
\label{projector}
\end{equation}
where $\delta_A^B$ is the Kronecker delta. Indeed,
\begin{equation}
(\mathcal P V)^A\,=\,V^A\,-\,\frac{X\cdot V}{X^2}\,X^A\,,\qquad X\cdot(\mathcal P V)=0 \,.
\label{projvect}
\end{equation}
More generally,
\begin{equation}
(\mathcal PT)_{A_1 \ldots  A_r}\,:=\, \mathcal P_{A_1}^{B_1}\, \ldots  \,\mathcal P_{A_r}^{B_r}\,T_{B_1 \ldots  B_r}\,,
\qquad X^{A_i}(\mathcal P T)_{A_1\ldots  A_i\ldots  A_r}=0
\label{raccourci}
\end{equation}

From now, all tensors will always be completely symmetric under the permutations of indices.
The leitmotiv of the present paper is to realize the space of symmetric tensor fields on $\mathcal{M} _d$ as a (sub)space of homogeneous symmetric tensor fields on ${\mathbb R}_0^{D}$. However, three distinct but equivalent realizations prove to be useful: either the ambient tensors are
\begin{enumerate}
	\item required to fulfill the condition $X^{A_1}\,T_{A_1\ldots A_r}(X) \,=\,0$, or
	\item projected by hand via the projector $\mathcal P$, or
	\item seen as equivalence classes of the relation 
\begin{equation}
T_{A_1\ldots A_r}\,\sim\, T_{A_1\ldots A_r}\,+\,X_{(A_1}U_{A_2\ldots A_r)}\,. 
\label{equrel}
\end{equation}
\end{enumerate}
Obviously, the first and second realization are equivalent to each other. The third realization is equivalent to the previous ones because the latter merely correspond to a particular choice of representative.

An important example is the induced metric, \textit{i.e.} the pull-back of the flat metric $\eta_{AB}$ which reads in intrinsic coordinates as
\begin{equation}
g_{\mu\nu} \,=\, \dfrac{\partial X^A}{\partial x^\mu}\,\dfrac{\partial X^B}{\partial x^\nu}\,\eta_{AB}\,=\,\dfrac{\partial X}{\partial x^\mu}\,\cdot\,\dfrac{\partial X}{\partial x^\nu}\,,
\label{defGab}
\end{equation}
which will be used to raise and lower the minuscule Greek indices.
The induced metric can be represented by the ambient tensor
\begin{equation}
G_{AB}\,=\,\mathcal P_A^C\,\mathcal P_B^D\,\eta_{CD}\,=\,\eta_{AB}\,-\,\dfrac{X_A X_B}{X^2}
\label{Gabexplicite}
\end{equation}
which is in the image of the projection operator $\cal P$ and obeys to the transversality condition $X^AG_{AB}=0$.
Notice that the ambient tensor $G_{AB}$ representing the induced metric $g_{\mu\nu}$ is in the same equivalence class as the ambient metric, $G_{AB} \,\sim\, \eta_{AB}$, as it should. Moreover, $G_A^B\,=\,\mathcal P_A^B$.

\subsection{Covariant derivatives}

The main technical difficulty in the ambient formulation is the translation of ambient partial derivatives $\partial_A$ in terms of intrinsic covariant derivatives.
In order to overcome this obstacle, a generating function performing the translation rule is provided in this subsection.

Let $\nabla_\mu$ be the covariant derivative corresponding to the Levi-Civita connection on the (pseudo) Riemannian manifold $\mathcal{M} _d$.
Its representative $\mathcal D$ in the ambient space ${\mathbb R}_0^{D}$ is the operator
\begin{equation}
\mathcal D\,=\,\mathcal P \,\circ\, \partial \, \circ \,\mathcal P \,.
\label{covder}
\end{equation}
A similar formulation of the covariant derivative in terms of the ambient partial derivative has been used in \cite{Metsaev}. For instance, the covariant derivative $\nabla_\mu v_\nu$ of a vector field $v_\mu$ on $\mathcal{M} _d\subset{\mathbb R}_0^{D}$ is represented in ambient space as
\begin{eqnarray}
\mathcal D_A V_B\,:=\,\mathcal P_A^C\,\mathcal P_B^D\,\partial_C(\mathcal P_D^E\,V_E) 
\label{deriveecovariantevecteur}
\end{eqnarray}
Geometrically, the definition (\ref{deriveecovariantevecteur}) means that the infinitesimal parallel transportation of a vector field $v_\mu$ on $\mathcal{M} _d$
can be performed in ambient space in three steps as follows: firstly, project on the tangent bundle $T\mathcal{M} _d$ its ambient representative $V_A$; secondly, infinitesimal parallel transport the resulting vector with respect to the ambient space metric; finally, project again the result on $T\mathcal{M} _d$.
Algebraically, the first step is the projection (\ref{projvect}), the second step is the mere partial derivation $\partial_C$, so that the third step indeed gives (\ref{deriveecovariantevecteur}).
One may prove algebraically that the definition (\ref{covder}) indeed implements the unique Levi-Civita connection $\nabla$ on $\mathcal{M} _d$ by checking that $\mathcal D$ verifies the following three axioms:
\begin{description}
	\item[-] Leibnitz rule: 
\begin{equation}
\mathcal D_A(\Phi_1 \,\Phi_2)\,=\,(\mathcal D_A\Phi_1)\Phi_2\,+\,\Phi_1\mathcal D_A\Phi_2\,\leftrightarrow\,\nabla_\mu(\phi_1\phi_2)=(\nabla_\mu\phi_1)\phi_2+\phi_1\nabla_\mu\phi_2\,,
\end{equation}
	\item[-] Metricity: $\mathcal D_A\,G_{BC}\,=\,0\,\leftrightarrow \,\nabla_\mu g_{\nu\rho}=0$\,,
	\item[-] Torsionlessness: $[\mathcal D_A,\mathcal D_B]\Phi\,=\,0\,\leftrightarrow\, [\nabla_\mu,\nabla_\nu]\phi=0$\,.
\end{description}
More concretely, the definition \eqref{covder} reads in components as  
\begin{eqnarray}
\label{explcovder}
\mathcal D_A T_{B_1 \ldots  B_r}\,:=\,\mathcal P_A^C\,\mathcal P_{B_1}^{D_1}\ldots  \mathcal P_{B_r}^{D_r}\,\partial_C\big(\mathcal P_{D_1}^{E_1}\, \ldots  \,\mathcal P_{D_r}^{E_r}\,T_{E_1 \ldots  E_r}\big)
\end{eqnarray}
where the definition \eqref{raccourci} of the projector $\mathcal P$ was used. Although this formula provides a nice way to compute covariant derivatives via mere partial derivations in ambient space, the intermediate projections quickly become cumbersome when the rank of the tensor or the number of derivatives becomes large. Fortunately, it is possible to obtain an explicit formula relating the usual partial derivatives in ambient space to the symmetrized covariant derivatives.

In order to express general formulae in compact terms, a standard trick is to contract every index with an auxiliary vector, say $P^A$\,: 
\begin{eqnarray}
\nonumber
T(X,P)&=& P^{A_1}\ldots P^{A_r}\,T_{A_1\ldots A_r}(X)\,,\\
\nonumber
 (P \cdot \partial)^n &=& P^{A_1}\ldots P^{A_n}\,\partial_{A_1}\ldots \partial_{A_n}\,,\\
\nonumber
(P \cdot \mathcal D)^n &=& P^{A_1}\ldots P^{A_n}\mathcal D_{(A_1}\ldots \mathcal D_{A_n)}\,,\\
\label{notat}
P^2 &=& P^{A}P^B\,\eta_{AB}\,.
\end{eqnarray}
One may express recursively the powers of ambient partial derivatives $\partial$ like polynomials of the covariant derivatives $\cal D$ and the flat metric: 
\begin{eqnarray}
(P\cdot\partial)^n\, T(X,P)\,=\sum \limits_{m=0}^{\left[ n/2 \right]}\,c_n^m\left(\dfrac{\,P^2}{\,X^2}\right)^m\,(P\cdot\mathcal D)^{n-2m}\,T(X,P)
\label{def}
\end{eqnarray}
where $[q]$ is the integer part of the rational number $q$ and the coefficients $c_n^m$ should be determined. The dependence of these coefficients $c_n^m$ on the homogeneity degree $k$ in $X$ and $r$ in $P$ will be left implicit for not overloading the formulae.
Notice that, by hypothesis,
$c_n^m\,=\,0$ when $m \,\geqslant (n+1)/2$
and $ c ^ 0_n \, = \, 1$ for all $n\in\mathbb N$.
The equation (\ref{def}) amounts to the following dictionary between ambient partial derivatives and intrinsic symmetrized covariant derivatives
\begin{eqnarray}
&&\partial_{(A_1}\ldots\partial_{A_n}T_{A_{n+1}\ldots A_{r+n})}\,\longleftrightarrow\,\nonumber\\
&&\quad\longleftrightarrow\,\sum \limits_{m=0}^{\left[ n/2 \right]}\,c_n^m\,\left(\frac{\,\pm1}{\,R^2}\right)^m g_{(\mu_1\mu_2}\ldots g_{\mu_{2m-1}\mu_{2m}}\,\nabla_{\mu_{2m+1}}\ldots\nabla_{\mu_n}\,t_{\mu_{n+1}\ldots\mu_{r+n})}
\label{dicopart}
\end{eqnarray}
In Appendix, one shows that the function (analytic near the origin) 
\begin{equation}
c(x,y;k-r) \,=\,\sum\limits_{n=0}^{\infty} \,\sum\limits_{m=0}^{\left[ n/2 \right]}\, \frac{1}{\,n!} \,\, c_n^m \, x^{n-2m}\,y^m\,=\,
(1\,+\,y)^{\frac{k-r}2}\,\exp\left( \frac{x}{\sqrt{y}}\,\arctan\sqrt{y}\right)
\label{cxy}
\end{equation}
is a generating function for the $c_n^m$ coefficients.
The non-vanishing coefficients for $m < (n+1)/2$ can be written explicitly by identifying the relevant coefficients in the power expansion (given for $r=0$):
\begin{eqnarray}
\nonumber
c_n^m=\,& &
\,\sum\limits_{i_{n\,-\,2m}=0}^{m}\,\,\sum\limits_{i_{n\,-\,2m\,-1}=0}^{i_{n\,-\,2m}}
...\sum\limits_{i_1=0}^{i_2}\,\frac{1}{(m-i_{n-2m})!} \left(\frac k2
\right)\left(\frac k2 -1\right)\ldots\left(\frac k2
-m+i_{n-2m}+1\right)\,\times  \\
\nonumber
&& \qquad\times\frac{n!}{(n-2m)!}\,\frac{\,(-1)^{i_{n\,-\,2m}}}{(2\,i_1\,+\,1)\,\,(2\,(i_2-i_1)+1)\,\,...\,\,(2\,(i_{n\,-\,2m}-i_{n\,-\,2m\,-1})\,+\,1)}\,.
\end{eqnarray}
For instance, the first coefficients are
\begin{center}
\begin{tabular}{lll}
$c_0^0\,=\,1\,,$ &   &   \tabularnewline
$c_1^0\,=\,1\,,$ &   &   \tabularnewline
$c_2^0\,=\,1\,,$ &  $c_2^1\,=\,k\,,$ &   \tabularnewline
$c_3^0\,=\,1\,,$ &  $c_3^1\,=\,3k\,-\,2\,,$ &   \tabularnewline
$c_4^0\,=\,1\,,$ &  $c_4^1\,=\,2\,(3k\,-\,4)\,,$ &  $c_4^2\,=\,3k\,(k\,-\,2)\,,$ \tabularnewline
... & ... & ...
\end{tabular}
 \end{center}
Therefore \eqref{dicopart} provides, for instance, the following translation rules:
\begin{eqnarray}
\nonumber
\partial_A\Phi &\longleftrightarrow& \nabla_\mu \, \phi \\
\nonumber
\partial_A\,\partial_B\Phi \,  &\longleftrightarrow&  \nabla_{(\mu}\,\nabla_{\nu)}\phi \,\pm\,\dfrac{k}{R^2}\, g_{\mu\nu} \,\phi \\
\nonumber
\partial_A\,\partial_B \partial_C\Phi &\longleftrightarrow& \nabla_{(\mu}\,\nabla_\nu \mathcal \nabla_{\rho)}\, \phi \,\pm\,\dfrac{3k-2}{R^2}\, g_{(\mu\nu} \,\nabla_{\rho)}\phi \\
\nonumber
\partial_A\,\partial_B \partial_C \partial_D\Phi &\longleftrightarrow& \nabla_{(\mu}\,\nabla_\nu \nabla_\rho \nabla_{\sigma)}\, \phi \, \,\pm\,\dfrac{2\,(3k-4)}{R^2}\, g_{(\mu\nu} \,\nabla_\rho \nabla_{\sigma)}\phi \\
&&+\,\dfrac{3k\,(k-2)}{R^4}\, g_{(\mu\nu}\,  g_{\rho\sigma)}\,\phi \\
\label{formulesderivee}
\nonumber
&\vdots&
\end{eqnarray}
Notice that a most compact and useful way to summarize \eqref{def} is as
\begin{equation}
T(X\,+\,t\,P\,,\,P)\,=\,c(t\,P\cdot\mathcal D\,,\,t^2\,P^2/X^2\,;\,k-r)\,T(X,P)\,,\qquad\forall t\,,
\label{Tx+p}
\end{equation}
as can be seen from the Taylor expansion of 
\begin{equation}
T(X+tP,P)=\exp(t\,P\cdot\partial)T(X,P)\,=\,\sum\limits_{n=0}^\infty\frac{t^n}{n!}\,(P\cdot\partial)^nT(X,P)
\end{equation}
in power series of $t$.

\subsection{Laplace-Beltrami operators}\label{Beltrami}

Combining the definitions \eqref{Gabexplicite} and \eqref{covder} of the last two subsections, one finds that the Laplace-Beltrami operator $\nabla^2\,=\,g^{\mu\nu}\,\nabla_\mu\,\nabla_\nu$ is represented in ambient space by $G^{AB}\,\mathcal D_A \, \mathcal D_B$. On rank-$r$ symmetric tensor fields, it acts as follows
\begin{eqnarray}
&&\nabla^2t_{\mu_1\ldots \mu_r}(x)\longleftrightarrow G^{BC}\,\mathcal D_B \, \mathcal D_CT_{A_1\ldots A_r}(X)\sim\nonumber\\
&&\qquad\qquad\qquad\sim\left[\partial^2\,-\,\dfrac{1}{X^2}\,(X\cdot\partial)\left(X\cdot\partial\,+\,D\,-\,2\,-\,r\right)\right]\,T_{A_1\ldots A_r}(X)
\label{nablacarre}
\end{eqnarray}
as can be checked explicitly. Therefore, the action of the ambient Laplace-Beltrami operator $\partial^2=\eta^{AB}\partial_A\partial_B$ on ambient symmetric tensor fields of homogeneity degree $k$ is translated in intrinsic components as follows
\begin{eqnarray}
\partial^2\,T_{A_1\ldots A_r}(X)\,\longleftrightarrow\,\left[\nabla^2\,\pm\,\dfrac{1}{R^2}\,k\left(k\,+\,d\,-\,1\,-\,r\right)\right]\,t_{\mu_1\ldots \mu_r}(x)\,.
\label{dalembertien}
\end{eqnarray}

For scalar fields ($r=0$), one recovers the standard formulae for the eigenvalues of the Laplace-Beltrami operator for the ``spherical'' harmonics in any dimension. 
In particular, when the number of timelike directions in the signature of the ambient metric $\eta$ is equal to:
\begin{itemize}
	\item Zero (Euclidean), the quadric $X^2=R^2$ is a hypersphere, $\mathcal M_d=S^d$, which can be seen as the Wick rotation of the de Sitter spacetime space $dS_d$. A textbook material on group theory is the fact that the genuine spherical harmonics with fixed homogeneity, \begin{equation}k_{S^d}=\ell\in\mathbb N\,,\end{equation} span unitary irreducible representations of $\mathfrak{o}\,(d+1)$. These spherical harmonics	
are the evaluation $\phi(x)$ on $S^d$ of homogeneous harmonic polynomials $\Phi(X)$ such that \eqref{homh},
\begin{eqnarray}
\partial^2\,\Phi(X)\,=\,0\,\longleftrightarrow\,\left[\Delta_{S^d}\,+\,\dfrac{1}{R^2}\,\ell\left(\ell\,+\,d\,-\,1\right)\right]\,\phi(x)\,=\,0\,.
\label{Laplacian}
\end{eqnarray}
  \item One (Lorentzian), the one-sheeted hyperboloid $X^2=+R^2$ is the de Sitter spacetime, $\mathcal M_d=dS_d$, while the two-sheeted hyperboloid
$X^2=-R^2$ is (two copies of) the hyperbolic space, $\mathcal M_d=H^d$. The unitary irreducible representations of $\mathfrak{o}\,(1,d)$ corresponding to massive scalar fields have been studied a while ago in \cite{Mickelsson:1972fh} and belong to the principal continuous series. They can be realized as the evaluation $\phi(x)$ on $dS_d$ of homogeneous harmonic functions $\Phi(X)$ of complex homogeneity degree $k_{dS_d}\in\mathbb C$ such that
\begin{equation}
\mathop{\mathrm{Re}}(k_{dS_d})\,=\,1\,-\,\dfrac D2 \,=\,\dfrac{1-d}{2}\,,\qquad \mathop{\mathrm{Im}}{(k_{dS_d})}\,=\,\mu\,,
\label{kds}
\end{equation}
where $\mu$ is a parameter with mass dimension.
This implies that the wave equation reads as
\begin{eqnarray}
\partial^2\,\Phi(X)\,=\,0\,\longleftrightarrow\,\left[\nabla^2_{dS_d}\,-\,\dfrac{1}{R^2}\, \left(\left(\dfrac{d\,-\,1}{2}\right)^2\,+\,\mu^2\right)\right]\,\phi(x)\,=\,0\,.
\end{eqnarray}
  \item Two (Conformal), the one-sheeted hyperboloid $X^2=-R^2$ is the anti de Sitter spacetime, $\mathcal M_d=AdS_d$, whose Wick rotation is the previous (two copies of the) hyperbolic space $H^d$. The lowest weight unitary irreducible representations of $\mathfrak{o}\,(2,d-1)$ corresponding to massive scalar fields on (the universal covering of) $AdS_d$ with energy bounded from below are well known (see \textit{e.g.} \cite{deWit:1999ui} for a nice review). They can be realized as the evaluation $\phi(x)$ on $AdS_d$ of homogeneous harmonic functions $\Phi(X)$ of real homogeneity degree $k_{AdS_d}\in\mathbb R$ such that
\begin{eqnarray}
\label{kads}
k_{AdS_d}\,=\,1\,-\,\dfrac D2 \,+\mu\,=\,\dfrac{1-d}{2}\,+\,\mu\,.
\end{eqnarray}
In any case, the corresponding wave equation is
\begin{eqnarray}
\partial^2\,\Phi(X)\,=\,0\,\longleftrightarrow\,\left[\nabla^2_{AdS_d}\,+\,\dfrac{1}{R^2}\, \left(\left(\dfrac{d\,-\,1}{2}\right)^2\,-\,\mu^2\right)\right]\,\phi(x)\,=\,0\,.
\end{eqnarray}
\end{itemize}

To summarize, the wave equation for a unitary massive scalar field on $(A)dS_d$ is
\begin{eqnarray}
\nabla^2_{(A)dS_d}\phi(x)\,=\,\dfrac{1}{R^2}\,\left( \pm\,\left(\dfrac{d\,-\,1}{2}\right)^2\,\, +\,\mu^2\right)\,\phi(x)\,=\, m^2\,\phi(x)\,,
\label{KGads}
\end{eqnarray}
where, as mentioned before the $\pm$ symbol refers to the corresponding equation $X^2=\pm R^2$.
Thus the unitary bound on the ``mass square'' (or, better, the eigenvalue of the quadratic Casimir operator of the isometry algebra) of a scalar field on $(A)dS_d$ is determined by the inequality
\begin{equation}
\big(mR\big)^2\,:=\,\pm\,\left(\dfrac{d\,-\,1}{2}\right)^2\,\, +\,\mu^2\,\geqslant\,\pm\,\left(\dfrac{d\,-\,1}{2}\right)^2\,,
\end{equation}
which reproduces the Breitenlohner-Freedman bound \cite{Breitenlohner:1982jf} in the $AdS_d$ case where (naive) ``tachyonic'' fields may be unitary and stable.
As one can see, the massive scalar field on  $AdS_d$ may be obtained as the analytic continuation of the massive scalar fields on $dS_d$ where  $\mu$ (and $R$) is replaced by $-i\mu$ (and $-iR$).

For later purpose, let us denote the ambient scalar field $\Phi^\dagger(X)$ as being the function on ${\mathbb R}_0^{D}$ whose homogeneity degree $k^\dagger_{(A)dS_d}$ is equal to $k_{(A)dS_d}$ up to the substitution of $\mu$ by $-\mu$ in \eqref{kds} or \eqref{kads} respectively, and whose evaluation on $(A)dS_d$ is equal to $\phi^*(y)$, \textit{i.e.}
\begin{equation}
\Phi^\dagger(X^A)\,=\,\Phi^\dagger(\rho,\,y^\mu)\,=\, \left(\dfrac{\rho}{R}\right)^{k^\dagger}\,\phi^*(y^\mu)\,.
\label{homhd}
\end{equation}
This homogeneous function $\Phi^\dagger(X)$ is also harmonic and the complex conjugate $\phi^*(x)$ satisfies to the same wave equation \eqref{KGads}.
A compact way to summarize the respective homogeneity degrees on $(A)dS_d$ is as follows:
\begin{eqnarray}
\nonumber
k_{(A)dS_d}&=&1\,-\,\dfrac D2 \,+\,\sqrt{\mp 1}\,\,\mu\,=\,\dfrac{1-d}{2}\,+\,\sqrt{\mp 1}\,\,\mu\,,
\\
\label{homogeneity}
k^\dagger_{(A)dS_d}&=&1\,-\,\dfrac D2 \,-\,\sqrt{\mp 1}\,\,\mu\,=\,\dfrac{1-d}{2}\,-\,\sqrt{\mp 1}\,\,\mu\,,
\end{eqnarray}
where, once again, the $\pm$ symbol refers to the corresponding equation $X^2=\pm R^2$.
Notice also the useful identities 
\begin{eqnarray}
\pm\big(mR\big)^2&=&-\,k_{(A)dS_d}(k_{(A)dS_d}\,+\,d\,-\,1)
\nonumber\\
&=&-\,k_{(A)dS_d}^\dagger(k_{(A)dS_d}^\dagger\,+\,d\,-\,1)\,\,,
\label{kkd}\\
&=&k^\dagger_{(A)dS_d} k_{(A)dS_d}\,.
\nonumber
\end{eqnarray}
In the AdS/CFT litterature, the opposite of $k_{AdS_d}$ and $k^\dagger_{AdS_d}$ are usually denoted by $\Delta_+$ and $\Delta_-$.

Various ambient/spacetime notations that have been introduced so far are summarized in a table at the very end of the paper.

\subsection{Klein-Gordon action}

The quadratic action of a complex massive scalar field on $(A)dS_d$ reads, modulo a boundary term, as 
\begin{eqnarray}
\label{KGact}
S_0[\phi]\,=\,-\,\frac12\int_{(A)dS_d} d^dx\,\sqrt{|\,g|}\,\left(\,g^{\mu\nu}\partial_\mu\phi^*(x)\partial_\nu\phi(x)\,+\,m^2\,|\,\phi(x)|^2\, \right).
\end{eqnarray}
It can be rewritten in the ambient formulation where the covariance under all isometries is manifest,
\begin{eqnarray}
&&S_0[\phi]\,=\, -\int_{{\mathbb R}^D_0}d^DX \,|\,X^2|^{\frac12}\,\delta(X^2\mp R^2)\times\nonumber\\
&&\qquad\qquad\qquad\times\left(\,G^{AB}\partial_A\Phi^{\dagger}(X)\,\partial_B\Phi(X)\,\pm\,\frac{(mR)^2}{X^2}\,\Phi^{\dagger}(X)\Phi(X)\,\right).
\label{KGamb}
\end{eqnarray}
In (pseudo) spherical coordinates, the volume form reads as 
\begin{equation}
\label{volumeform}
d^DX \,=\, d\rho\,\left(\frac{\rho}{R}\right)^d\,d^dx\,\sqrt{|\,g(x)|}\,,
\end{equation}
In order to check the equality \eqref{KGamb}, one should rewrite the integral over ${\mathbb R}^D_0$ in (pseudo) spherical coordinates, insert the homogeneity conditions \eqref{homh} and \eqref{homhd} as well as the following relation on the Dirac delta function, \begin{equation}|\,X^2|^{\frac12}\,\delta(X^2\mp R^2)\,=\,\rho\,\delta(\rho^2\mp R^2)\,=\,\frac{\rho}{|\rho+R|}\,\delta(\rho\,-\,R)\,=\,\frac12\,\delta(\rho\,-\,R)\,,\end{equation} and, finally, integrate over the radial coordinate $\rho$ from zero to infinity.

There is also an alternative way to obtain the spacetime integral \eqref{KGact} in a form where the covariance under all isometries is manifest:
along the lines of the radial dimensional reduction from massless to massive fields and from flat to curved spacetimes \cite{Biswas:2002nk}, one may
instead remove the Dirac delta $\delta(\rho-R)$ in the integral over the ambient space.
With the help of \eqref{kkd} and
\begin{equation}
G^{AB}\partial_A\Phi^{\dagger}(X)\,\partial_B\Phi(X)\,=\,\eta^{AB}\partial_A\Phi^{\dagger}(X)\,\partial_B\Phi(X)\,-\,\frac1{X^2}\,(X\cdot\partial)\Phi^{\dagger}(X)\,(X\cdot\partial)\Phi(X)\,,
\end{equation}
together with \eqref{volumeform}, one can show that
\begin{eqnarray}
\nonumber
S_0[\Phi]&:=&-\frac12\int_{{\mathbb R}^D_0}d^DX\,
\eta^{AB}\partial_A\Phi^{\dagger}(X)\,\partial_B\Phi(X)\,
\\
&=&-\frac12\int_{{\mathbb R}^D_0}d^DX\,
\left(\,G^{AB}\partial_A\Phi^{\dagger}(X)\,\partial_B\Phi(X)\,\pm\,\frac{(mR)^2}{X^2}\,\Phi^{\dagger}(X)\Phi(X)\,\right)
\label{radimred}\\
\nonumber
&=&R\int_0^\infty dz\,\times\,
S_0[\phi]
\end{eqnarray}
where the integral over $z$ on the right-hand-side is simply a constant factor (albeit infinite)
Remember that $z=\log(\rho/R)$ and $(\rho/R)^k=\exp(k\,z)$.
The analogy of \eqref{radimred} with a dimensional reduction along a (non-compact) direction further justified the choice of terminology ``radial dimensional reduction'' in \cite{Biswas:2002nk}. This interpretation is somewhat more natural in $dS_d$ where the radial direction is spacelike (though non-compact) as it should and where $\Phi^\dagger$ is simply the complex conjugate of $\Phi$. In this analogy, the parameter $\mu$ plays the usual role of the mass for the Fourier factor $\exp(i\,\mu\,z)$. The basis of the radial dimensional reduction technique is the observation that, since the kinetic operator for massless fields on flat spacetime is scale invariant, the homogeneity condition on the fields is a consistent ansatz. Moreover, the homogeneity degree must be chosen such that the action on the flat ambient space is also scale invariant.

\subsection{Noether method}\label{Noeth}

The ambient formalism developed above should also be applied to the whole content of the section \ref{Noethermethod}.
In this subsequent, one introduces various definitions dedicated to an ambient reformulation of Section \ref{Noethermethod}, preparing the ground for the next two sections.

The \textit{ambient representative of a symmetric conserved current} of rank $r\geqslant 1$, say $j^{\mu_1\ldots\,\mu_r}$, is an equivalence class $J^{\,A_1\ldots\,A_r}\,\sim\, J^{A_1\ldots A_r}\,+\,X^{(A_1}U^{A_2\ldots A_r)}$ of real contravariant homogeneous symmetric tensor fields on ${\mathbb R}_0^{D}$  
of homogeneity degree equal to $2-D-r$
where one of the representative obeys to the strict conservation law 
\begin{equation}
\partial_{A_1}J^{A_1\ldots\,A_r}(X)\approx 0\,.
\label{strictconservationlaws}
\end{equation}
The homogeneity degree, 
\begin{equation}
(X^A \partial_A\,+\,D\,-\,2\,+\,r)J^{A_1\ldots\,A_r}(X)= 0\,,
\end{equation}
is chosen such that the equation \eqref{strictconservationlaws} is preserved by the equivalence relation, as can be checked directly and as will be shown later in a more economical way.
This property implies the covariant conservation law
\begin{equation}
\mathcal D_{A_1}J^{A_1\ldots\,A_r}(X)\approx 0\,.
\label{covconservationlaws}
\end{equation}
corresponding to \eqref{conservationlaws}, even though the representative $J^{A_1\ldots\,A_r}(X)$ satisfying \eqref{strictconservationlaws} may not be tangent.
An \textit{ambient generating function of conserved currents} is an equivalence class
\begin{equation}
J(X,P)\,\sim\,J(X,P)\,+\,(X\cdot P) U(X,P)\,\Longleftrightarrow\, J^{A_1\ldots A_r}\,\sim\, J^{A_1\ldots A_r}\,+\,r\,X^{(A_1}U^{A_2\ldots A_r)}\,.
\end{equation}
of real functions on the phase space $T^*{\mathbb R}_0^{D}$ which are (i) formal power series in the momenta, (ii) such that
\begin{equation}
\left(X^A \frac{\partial}{\partial X^A}\,+\,P_A\frac{\partial}{\partial P_A}\,+\,D\,-\,2\right)J(X,P)\,=\, 0\,,
\label{homXP}
\end{equation}
\begin{equation}
\left(X^A \frac{\partial}{\partial X^A}\,+\,P_A\frac{\partial}{\partial P_A}\,+\,D\,\right)U(X,P)\,=\, 0\,,
\label{hoomXP}
\end{equation}
and (iii) where one of the representatives obeys to
\begin{equation}
\left(\frac{\partial}{\partial X^A}\frac{\partial}{\partial P_A}\right)\,J(X,P)\approx 0\,.
\label{ambientcons}
\end{equation}
The commutation relation
\begin{equation}
\left[\frac{\partial}{\partial X^A}\frac{\partial}{\partial P_A}\,,\,X^BP_B\right]\,=\,X^A \frac{\partial}{\partial X^A}\,+\,P_A\frac{\partial}{\partial P_A}\,+\,D
\end{equation}
implies that, provided the homogeneity condition \eqref{homXP} is satisfied (which is consistent with the radial reduction ansatz), the ambient divergence is well defined on equivalence classes of currents, \textit{i.e.}
\begin{equation}
J_1\sim J_2\,\Longrightarrow\,\left(\dfrac{\partial}{\partial X^A}\dfrac{\partial}{\partial P_A}\right)J_1\,\sim\,\left(\dfrac{\partial}{\partial X^A}\dfrac{\partial}{\partial P_A}\right)J_2\,,
\label{equivcldiv}
\end{equation}
because $[\partial_X\cdot\partial_P\,,\,X\cdot P]U=0$ due to \eqref{hoomXP}.
Therefore, the current is covariantly divergenceless
\begin{equation}
\left(\mathcal D_A\dfrac{\partial}{\partial P_A}\right)J(X,P)\,\approx\,0
\label{ambientcovcons}
\end{equation}
when \eqref{homXP} holds since \eqref{ambientcons} and \eqref{equivcldiv} imply \eqref{ambientcovcons}.
Thus all the coefficients of order $r\geqslant 1$ in the power expansion of the generating function
\begin{equation}
J(X,P)\,=\,\sum\limits_{r\geqslant 0}\frac{1}{r!}\,J^{A_1\ldots\,A_r}(X)\,P_{A_1}\ldots  P_{A_r}
\label{J}
\end{equation}
are all ambient representative of conserved currents by means of (\ref{ambientcovcons}).

The \textit{ambient representative of a symmetric tensor gauge field} of rank $r\geqslant 1$, say $h_{\mu_1\ldots \mu_r}(x)$, is a real covariant homogeneous symmetric tangent tensor field $H_{A_1\ldots A_r}(X)$ on ${\mathbb R}_0^{D}$ 
of homogeneity degree equal to $r-2$
whose gauge transformations are of the form
\begin{equation}
\delta_\epsilon H_{A_1\ldots A_r}(X)\,=\,r\,\partial_{(A_1}\epsilon_{A_2\ldots A_r)}(X)\,+\,{\cal O}(H)\,=\,r\,\mathcal D_{(A_1}\epsilon_{A_2\ldots A_r)}(X)\,+\,{\cal O}(H)\,,
\end{equation}
where the gauge parameter $\epsilon_{A_1\ldots A_{r-1}}(X)$ is a covariant homogeneous symmetric tangent tensor field on ${\mathbb R}_0^{D}$ of rank $r-1$ and of homogeneity degree $r-1$.
The homogeneity degrees are such that the symmetrized gradient of $\epsilon$ is tangent, as can be checked by direct computation, so that
$\partial_{(A_1}\epsilon_{A_2\ldots A_r)}(X)=\mathcal D_{(A_1}\epsilon_{A_2\ldots A_r)}(X)$.
An \textit{ambient generating function of gauge fields} is a real function $H(X,V)$ on the configuration space $T{\mathbb R}_0^{D}$ (i) which is a formal power series in the velocities, (ii) such that 
\begin{equation}
\left(X^A \frac{\partial}{\partial X^A}\,-\,V^A\frac{\partial}{\partial V^A}\,+\,2\right)H(X,V)\,=\, 0\,,\qquad
\left(X^A \frac{\partial}{\partial V^A}\right)H(X,V)\,=\, 0\,,
\label{hoXP}
\end{equation}
and (iii) whose gauge transformations are
\begin{equation}
\delta_\epsilon H(X,V)\,=\,\left(V^A\partial_A\right)\,\epsilon(X,V)\,+\,{\cal O}(H)\,
=\,\left(V^A\mathcal D_A\right)\,\epsilon(X,V)\,+\,{\cal O}(H)\,,
\label{ambFronsdalgtransfo}
\end{equation}
where $\epsilon(X,V)$ is a formal power series in the velocities such that
\begin{equation}
\left(X^A \frac{\partial}{\partial X^A}\,-\,V^A\frac{\partial}{\partial V^A}\,\right)\epsilon(X,V)\,=\, 0\,,\qquad
\left(X^A \frac{\partial}{\partial V^A}\right)\epsilon(X,V)\,=\, 0\,.
\label{eoXP}
\end{equation}
The commutation relation
\begin{equation}
\left[X^A\frac{\partial}{\partial V^A}\,,\,V^B\dfrac{\partial}{\partial X^B}\right]\,=\,X^A \frac{\partial}{\partial X^A}\,-\,V^A\frac{\partial}{\partial V^A}\,,
\end{equation}
implies that, provided \eqref{eoXP} is satisfied, then $(X\cdot\partial_V)\delta_\epsilon H(X,V)=\mathcal O(H)$.
The coefficients of order $r\geqslant 1$ in the power expansion of the generating function
\begin{equation}
H(X,V)\,=\,\sum\limits_{r\geqslant 0}\frac{1}{r!}\,H_{A_1\ldots\,A_r}(x)\,V^{A_1}\ldots V^{A_r}
\label{H}
\end{equation}
are all ambient representatives of symmetric tensor gauge fields due to (\ref{Fronsdalgtransfo}) with 
\begin{equation}
\epsilon(X,V)\,=\,\sum\limits_{t\geqslant 0}\frac{1}{t!}\,\epsilon_{A_1\ldots\,A_t}(X)\,V^{A_1}\ldots V^{A_t}\,.
\end{equation}

The non-degenerate bilinear pairing \eqref{pairing} between smooth functions $h(x,v)$ and $j(x,p)$ on the configuration and phase spaces respectively,
can be written in terms of the ambient representatives in a similar fashion to \eqref{KGamb}:
\begin{eqnarray}
&&\ll h\,\|\,j \gg\,:=\,\nonumber\\
&&\quad:=\,2\int_{{\mathbb R}^D_0}d^DX \,|\,X^2|^{\frac12}\,\delta(X^2\mp R^2) \,\left.\exp\left(\,\frac{{\partial}}{\partial V^A}\,\frac{{\partial}}{\partial P_A}\right)H(X,V)\,J(X,P)\,\right|_{V=P=0}
\label{pairingama}\\
\nonumber
&&\quad=\,2\sum\limits_{r\geqslant 0}\,\frac{1}{r!}\,\int_{{\mathbb R}^D_0}d^DX \,|\,X^2|^{\frac12}\,\delta(X^2\mp R^2)\,
\,H_{A_1\ldots A_r}(X)\,J^{\,A_1\ldots A_r}(X)\,.
\end{eqnarray}
Another option is to follow the philosophy of the radial dimensional reduction, as in \eqref{radimred}, 
\begin{eqnarray}
\ll H\,\|\,J \gg\,&:=&\int_{{\mathbb R}^D_0} d^DX\,\left.\exp\left(\,\frac{{\partial}}{\partial V^A}\,\frac{{\partial}}{\partial P_A}\right)H(X,V)\,J(X,P)\,\right|_{V=P=0}\label{pairingamb}\\
&=&\sum\limits_{r\geqslant 0}\,\frac{1}{r!}\,\int_{{\mathbb R}^D_0}d^DX \,
\,H_{A_1\ldots A_r}(X)\,J^{\,A_1\ldots A_r}(X)\,\nonumber\\
&=&R\int_0^\infty dz\,\ll h\,\|\,j \gg\nonumber
\end{eqnarray}
where the integrand of the integral 
over ${\mathbb R}^D_0$  on the second line is of homogeneity degree equal to $-D$ as it should.
This shows that if the conserved currents of the matter fields on a flat spacetime define ambient representatives with the right properties (such as their degree of homogeneity) then the radial dimensional reduction of the Noether {interaction} can be applied:
\begin{eqnarray}
S_1[\Phi,H]&:=&\ll H\,\|\,J \gg
\nonumber\\
&=&R\int_0^\infty dz\,\times S_1[\phi,h]
\label{radimredamb}
\end{eqnarray}

The \textit{ambient representative of a Killing tensor field} of rank $r-1\geqslant 0$ on $\mathcal{M} _d$ is a covariant homogeneous symmetric tangent tensor field $\overline{\epsilon}_{A_1\ldots A_{r-1}}(X)$ on ${\mathbb R}_0^{D}$ of degree $r-1$
solution of the generalized Killing equation
\begin{equation}
\partial_{(A_1}\overline{\epsilon}_{A_2\ldots A_r)}(X)=0\,.
\end{equation}
An \textit{ambient generating function of Killing fields} is a function $\overline{\epsilon}(X,V)$ on the configuration space $T{\mathbb R}_0^{D}$ which is a formal power series in $X^{[A}V^{B]}:=X^AV^B-X^BV^A$.
Then the coefficients in the power series
\begin{equation}
\overline{\epsilon}(X,V)\,=\,\overline{\epsilon}\left(X^{[A}V^{B]}\right)\,=\,\sum\limits_{t\geqslant 0}\frac{1}{t!}\,\overline{\epsilon}_{A_1\ldots A_t}(X)\,V^{A_1}\ldots V^{A_t}
\end{equation}
provide the most general ambient representatives of Killing tensor fields on $\mathcal{M} _d$
(see \textit{e.g.} \cite{Bekaert:2005ka,Bekaert:2008sa} for reviews and refs therein).

\vspace{1mm}
In the next two sections, these general facts will be applied to the case of a free complex scalar field.

\section{Conserved currents}\label{conscur}

The generating function of conserved currents on the flat ambient space \cite{Bekaert:2009ud} is equal to
\begin{eqnarray}
J(X,P) \,=\, \Phi^{\dagger}\left(X\,-\,i\,P\right)\,\Phi\left(X\,+\,i\,P\right)
\label{gencourant}
\end{eqnarray}
so that the corresponding ambient conserved currents take the explicit form
\begin{eqnarray}
J_{A_1 \ldots  A_r}(X) \,
&=&\,i^{\,r} \sum\limits_{s=0}^r\, (-1)^s \, \dbinom{r}{s}  \, \partial_{(A_1} \ldots  \,\partial_{A_s}\Phi^{\dagger}(X) \, \partial_{A_{s+1}} \ldots \, \partial_{A_r)}\Phi(X)
\nonumber \\
&=& i^{\,r}\,\Phi^{\dagger}(X)\,\overleftrightarrow{\partial_{A_1}} \ldots  \overleftrightarrow{\partial_{A_r}} \Phi(X)
\label{courantambiant}
\end{eqnarray}
where the usual double arrow $\overleftrightarrow{\partial}$ is defined by
\begin{equation}
\Phi\overleftrightarrow{\partial_A}\Psi\,:=\,\Phi({\partial_A}\Psi)\,-\,({\partial_A}\Phi)\Psi\,.
\end{equation}
These flat space currents (\ref{courantambiant}) are proportional to the ones introduced by Berends, Burgers and van Dam a long time ago \cite{Berends:1985xx}. Various explicit sets of (conformal) conserved currents on Minkowski spacetime were provided in \cite{current}.
The symmetric conserved current (\ref{courantambiant}) of rank $r$ is bilinear in the scalar field and contains exactly $r$ derivatives.
The currents of any rank are real thus, if the scalar field is real then the odd rank currents are absent due to the factor in front of (\ref{courantambiant}).
The generating function \eqref{gencourant} verifies 
\eqref{ambientcons}
when the ambient scalar field $\Phi$ obeys to the Klein-Gordon equation.
Although the ambient currents \eqref{courantambiant} are not tangent in general, they obey to \eqref{homXP} for homogeneous ambient scalar fields corresponding to massive scalar fields on $(A)dS_d$, since \eqref{homogeneity} implies 
\begin{equation}
k_{(A)dS_d}+k_{(A)dS_d}^\dagger\,=\,2-D\,,
\end{equation}
and therefore the previous equation \eqref{ambientcons}
is equivalent to the covariant conservation law \eqref{conservationlaws}. In other words, the radial dimensional reduction of the {cubic Noether interaction} is valid precisely for the mass-square domain of unitarity in $(A)dS_d$.

The main drawback of the explicit expressions \eqref{courantambiant} for the conserved currents
is that it is written in terms of ambient partial derivatives instead of covariant derivatives,
but the ambient generating function \eqref{gencourant} of $(A)dS_d$ conserved currents can be written very explicitly
in terms of \eqref{cxy} with the help of \eqref{Tx+p}
\begin{eqnarray}
\nonumber
J(X,P) &=& c\left(-\,i\,P\cdot\mathcal D\,, -\frac{P^2}{X^2}\,;\,k_{(A)dS_d}^\dagger\right)\,\,\Phi^{\dagger}(X) \,\, c\left(i\,P\cdot\mathcal D\,, -\frac{P^2}{X^2}\,;\,k_{(A)dS_d}\right) \, \Phi(X)\\
\nonumber
&=& \Phi^{\dagger}(X) \,c\left(-\,i\,P\cdot\overleftarrow{\mathcal D}\,, -\frac{P^2}{X^2}\,;\,k_{(A)dS_d}^\dagger\right) 
c\left(i\,P\cdot\overrightarrow{\mathcal D}\,, -\frac{P^2}{X^2}\,;\,k_{(A)dS_d}\right) \, \Phi(X)\\
\label{ambgen}
&=& \Phi^{\dagger}(X) \,c\left(i\,P\cdot\overleftrightarrow{\mathcal D}\,, -\frac{P^2}{X^2}\,;\,2-D\right) \, \Phi(X)
\end{eqnarray}
where the property $c(x_1,y\,;k_1)c(x_2,y\,;k_2)=c(x_1+x_2,y\,;k_1+k_2)$ and \eqref{homogeneity} were used.
The ambient generating function \eqref{ambgen} translates into the following generating function of conserved currents
\begin{equation}
j\,(x,p) \,=\, \phi^*(x)\,\,c\left(i\,p^\mu\overleftrightarrow\nabla_\mu\,, \mp\frac{g_{\mu\nu}p^\mu p^\nu}{R^2}\,;\,1-d\right) \, \phi(x)
\end{equation}
The flat limit is recovered for $R^2\rightarrow\infty$ since $c(x,y)\sim \exp x$ when $y\rightarrow 0$.
Due to \eqref{cxy}, the development \eqref{j} of this generating function gives the following conserved current of rank $r$,
\begin{equation}
j_{\mu_1 \ldots \mu_r}(x)\,=\,i^{\,r}\sum\limits_{m=0}^{\left[ r/2 \right]}\,\left(\frac{\,\mp1}{\,R^2}\right)^m c_r^m\,
\, g_{(\mu_1\mu_2}\ldots g_{\mu_{2m-1}\mu_{2m}}\,
\phi^*(x)\overleftrightarrow\nabla_{\mu_{2m+1}}\ldots \overleftrightarrow\nabla_{\mu_r)}\phi(x)\,,
\label{courantgeneral}
\end{equation}
where the coefficients $c_r^m$ correspond to $k=1-d$.
It is possible to compute numerically these coefficients $c_r^m$, the covariant derivatives \eqref{formulesderivee} and these currents from \eqref{courantgeneral} whatever the rank.
For example, we therefore find the first currents, which are all preserved by construction and which was also verified explicitly, calculated classically:
\begin{eqnarray}
\nonumber
j_{\mu}&=&i\,\, \phi^*\overleftrightarrow\nabla_\mu\phi\\
\nonumber
j_{\mu\nu}&=&-\, \phi^* \overleftrightarrow\nabla_\mu\overleftrightarrow\nabla_\nu\,\phi\,\pm\,\dfrac{1\,-\,d}{R^2}\,g_{\mu\nu}\,\phi^*\,\phi \\
\nonumber
j_{\mu\nu\rho}&=& -\,i\,\,\phi^* \overleftrightarrow\nabla_{(\mu}\overleftrightarrow\nabla_\nu\overleftrightarrow\nabla_{\rho)}\,\phi\,\pm\,i\,\dfrac{1\,-\,3d}{R^2}\,g_{(\mu\nu}\,\phi^*\,\overleftrightarrow\nabla_{\rho)}\,\phi\\
\nonumber
j_{\mu\nu\rho\sigma}&=& \phi^* \overleftrightarrow\nabla_{(\mu}\overleftrightarrow\nabla_\nu\overleftrightarrow\nabla_{\rho}\overleftrightarrow\nabla_{\sigma)}\,\phi\,\pm\,2\,\dfrac{1\,+\,3d}{R^2}\,g_{(\mu\nu}\,\phi^*\,\overleftrightarrow\nabla_\rho\overleftrightarrow\nabla_{\sigma)}\phi\\
\nonumber
&&+\,3\,\dfrac{d^2\,-\,1}{R^4}\,g_{(\mu\nu}\,g_{\rho\sigma)}\phi^*\,\phi \\
\nonumber
&\vdots&
\end{eqnarray}

Similar conserved currents on constant-curvature spaces were described in \cite{Prokushkin:1999ke,Ruhl,Fotopoulos} but the present results are somewhat more general: firstly, the currents (\ref{courantambiant}) are conserved for any free massive scalar field in any dimension, while only the conformal scalar (\textit{i.e.} the singleton) was considered in \cite{Ruhl} and $AdS_3$ was the background spacetime in \cite{Prokushkin:1999ke}; secondly, the explicit expression of the currents is known at all orders in the scalar curvature, while only the first order correction to the flat expression was provided in \cite{Ruhl}; thirdly, the currents (\ref{courantambiant}) are conserved on-shell in the usual sense of \eqref{conservationlaws} while the ones of \cite{Fotopoulos} obey to the weaker conservation law introduced by Fronsdal \cite{Fronsdal:1978vb}. Of course, strictly speaking the third comment should not be understood as a loss of generality in the previous results of \cite{Fotopoulos,Fotopoulos:2010nj}. We simply want to stress that usual conservation laws for the currents is a desirable property because it allows a uniform treatment of (ir)reducible gauge fields, \textit{e.g.} of triplet and Fronsdal fields, and it might also simplify the analysis of current exchange amplitudes.

\section{Noether interactions}\label{Noethercoupl}

As explained in the previous section, the function \eqref{gencourant} obeys to all properties for an ambient generating function of conserved currents, as defined in Subsection \ref{Noeth}. Therefore, the radial dimensional reduction of the corresponding ambient Noether {interaction} \eqref{pairingamb} is consistent and provides the Noether {interaction} \eqref{pairingsum} on $(A)dS_d$ where the conserved currents are given by \eqref{courantgeneral}. 
An important consequence of this fact is that one can import from flat spacetime all relationships (observed in \cite{Bekaert:2009ud}) between the Noether interactions of a complex scalar field with a collection of symmetric tensor gauge fields. In other words, the consistency of the radial dimensional reduction implies that one can induce the Weyl/Wigner technology used in \cite{Bekaert:2009ud} from the flat ambient space ${\mathbb R}^D_0$ onto the spacetime $(A)dS_d$. In this way, one reproduces the ambient approach to the Weyl/Wigner quantization of the cotangent bundle $T^*\mathcal M_d$ of a constant-curvature manifold, which was first introduced in the seminal papers on deformation quantisation with humor under the name ``a star product is born'' \cite{Bayen:1977ha}. The relevance of the latter approach to higher-spin gauge theory on (anti) de Sitter spacetime was argued in \cite{Bekaert:2008sa}.

\subsection{Symbol calculus}\label{symbolcalc}

Let us become more explicit. To start with, since ${\mathbb R}^D_0$ and $(A)dS_d$ are endowed with a metric, their respective tangent and cotangent spaces may be identified and thus one can identify ``momenta'' with ``velocities'', \textit{e.g.}
\begin{equation}P_A\,=\, \eta_{AB}\,V^B \qquad \text{and} \qquad p_{\mu}\,=\,g_{\mu\nu}v^\nu\,.\end{equation}
The ambient generating function of gauge fields $H(X,P)$ is now a real function on $T^*{\mathbb R}_0^{D}$ such that 
\begin{equation}
\left(X^A \frac{\partial}{\partial X^A}\,-\,P_A\frac{\partial}{\partial P_A}\,+\,2\right)H(X,P)\,=\, 0\,,\qquad
\left(X\cdot \frac{\partial}{\partial P}\right)H(X,P)\,=\, 0\,,
\label{hooXP}
\end{equation}
and whose gauge transformations are
\begin{equation}
\delta_\epsilon H(X,P)\,=\,\left(P\cdot \frac{\partial}{\partial X}\right)\,\epsilon(X,P)\,+\,{\cal O}(H)\,,
\label{ambFronsdalgtransfoo}
\end{equation}
where $\epsilon(X,P)$ is such that
\begin{equation}
\left(X^A \frac{\partial}{\partial X^A}\,-\,P_A\frac{\partial}{\partial P_A}\,\right)\epsilon(X,P)\,=\, 0\,,\qquad
\left(X\cdot \frac{\partial}{\partial P}\right)\epsilon(X,P)\,=\, 0\,.
\label{eooXP}
\end{equation}
The cotangent bundle $T^*\mathcal M_d$ can be seen as the sub-bundle of ${\mathbb R}^D_0$ defined by the quadric definition $X^2=\pm R^2$ together with the transversality condition $X^AP_A=0$. As symplectic manifolds, this embedding corresponds to a reduction with respect to the previous two constraints.

The ambient Moyal product of two smooth functions on $T^*{\mathbb R}^D_0$ is defined  by 
\begin{equation}
\epsilon_1(X,P) \star \epsilon_2(X,P) \, = \, \epsilon_1(X,P) \, \exp\left(\frac 12\,\overleftarrow{\frac{\partial}{\partial P_A}} \wedge \overrightarrow{\frac{\partial}{\partial X^A}}\right) \, \epsilon_2(X,P)
\label{Moyal}
\end{equation}
where $\wedge$ stands for the antisymmetric product.  
The conditions \eqref{eooXP} on $\epsilon(X,P)$ are equivalent to
\begin{equation}
[X \cdot P \, \stackrel{\bigstar}{,} \,\epsilon(X,P)] \,=\,0\,,\qquad
[X^2 \, \stackrel{\bigstar}{,} \, \epsilon(X,P)] \,=\,0\,.
\label{conde}
\end{equation}
where 
\begin{eqnarray}
\nonumber
[\epsilon_1(X,P) \, \stackrel{\bigstar}{,} \epsilon_2(X,P)\,]&:=&\epsilon_1(X,P) \star \epsilon_2(X,P)-\epsilon_2(X,P) \star \epsilon_1(X,P) \\
& =& \epsilon_1(X,P) \, 2\sinh\left(\frac 12\, \overleftarrow{\frac{\partial}{\partial P_A}} \wedge \overrightarrow{\frac{\partial}{\partial X^A}}\right) \, \epsilon_2(X,P)
\end{eqnarray}
denotes the ambient Moyal commutator.
The conditions \eqref{conde} expressed in terms of the Hermitian operator $\hat\epsilon$ the Weyl symbol of which is $\epsilon(X,P)$ state that this operator preserves the homogeneity degree and commutes with $X^2$.
The evaluation $\varepsilon(x,p)$ of the ambient representatives $\epsilon(X,P)$ provides an isomorphism between the space of smooth functions on $T^*\mathcal M_d$ and the (sub)space of smooth functions on $T^*{\mathbb R}^D_0$ which are subject to \eqref{conde}.
Moreover, the space of symbols obeying to \eqref{conde} is a subalgebra of the ambient Weyl algebra.
Therefore the pull-back of the Moyal product on $T^*{\mathbb R}^D_0$ induces a star product $*$ on the cotangent bundle $T^*\mathcal M_d$ such that the former isomorphism becomes an isomorphism of associative algebras, as pointed out by Bayen, Flato, Fronsdal, Lichnerowicz and Sternheimer in
\cite{Bayen:1977ha}. 
Notice that the Lie algebra of smooth functions on $T^*\mathcal M_d$ endowed with the corresponding star commutator $[\, \, \stackrel{*}{,} \,\,]$ is isomorphic to the Lie algebra of Hermitian (pseudo)differential operators on $\mathcal M_d$.
The adjoint action of this Lie algebra preserves the space of Weyl symbols such that \eqref{hooXP} and the gauge transformations \eqref{ambFronsdalgtransfoo} can be written as
\begin{equation}
\delta_\epsilon H(X,P)\,=\,\frac12\,[P^2 \, \stackrel{\bigstar}{,} \, \epsilon(X,P)]\,+\,{\cal O}(H)\,.
\label{ambgtransfoo}
\end{equation}
The ambient generating functions of Killing fields $\overline\epsilon(X,P)$ are Weyl symbols commuting with the three constraints $X^2$, $X\cdot P$ and $P^2$ which generate an $\mathfrak{sp}(2)$ algebra. The Lie (sub)algebra of such symbols is the off-shell higher-spin algebra of Vasiliev (see \textit{e.g.} \cite{Vasiliev:2004qz} for reviews).

\subsection{Cubic vertex}

Using the bra-ket notation for the scalar field $\Phi(X)=\bk X \Phi$ and $\Phi^\dagger(X)=\bk \Phi X$, the ambient generating function
$J(X,P)$ of currents \eqref{gencourant} is the (analytic continuation of the) Fourier transform over momentum space of the Wigner
function associated to the density operator $\ket\Phi\!\bra\Phi$ and the ambient Noether {interaction} \eqref{pairingamb} can be rewritten in a compact form as \cite{Bekaert:2009ud}
\begin{eqnarray}
S_1[\Phi,H]\,=\,\,\ll H\,\|\,J \gg\,\, =\, \bra\Phi \hat{H}\ket \Phi
\end{eqnarray}
where $H(X,P)$ is the Weyl symbol of the operator $\hat{H}$.

The ambient Klein-Gordon action \eqref{radimred} can be rewritten along the same lines as
\begin{eqnarray}
S_0[\Phi]\,=\,\bra\Phi \hat{H}_0\ket \Phi
\end{eqnarray}
where the operator $\hat{H}_0$ is defined by 
\begin{eqnarray}
\hat{H}_0:=\frac12\left[\partial^2\,-\,\dfrac{1}{X^2}\,(X\cdot\partial)\left(X\cdot\partial\,+\,D\,-\,2\right)\mp\,\frac{(mR)^2}{X^2}\right]
\end{eqnarray}
and is the ambient representative of the kinetic operator $\frac12(\nabla^2_{AdS_d}\,-\,m^2)$. It has Weyl symbol equal to 
\begin{equation}
H_0(X,P):=\frac12\left(G^{AB}P_AP_B\mp\,\frac{(mR)^2}{X^2}\right)=\frac12\left(P^2-\frac{(X\cdot P)^2}{X^2} \mp\,\frac{(mR)^2}{X^2}\right)
\end{equation}
where the transverse inverse metric $G^{AB}:=\eta^{AB}-X^AX^B/X^2$ is the ambient representative of the inverse metric $g^{\mu\nu}$ on $(A)dS_d\,$. Remark that the function $H_0(X,P)$ also obeys to \eqref{hooXP}.

Therefore the sum
\begin{eqnarray}
S_0[\Phi] \,+\, S_1[\Phi,H]\,=\,\,\bra\Phi \hat{H}_0+\hat{H}\ket \Phi
\end{eqnarray}
is manifestly invariant under the following action of the group of unitary operators on $(A)dS_d$:
\begin{equation}
    \ket\Phi\ \longrightarrow\ \hat{U}\ket\Phi\,,
    \qquad
    \hat{H}_0+\hat{H}\ \longrightarrow\ \hat{U}\ (\hat{H}_0+\hat{H})\ \hat{U}^{-1}\,,
    \label{finite}
\end{equation}
where the unitary operator $\hat{U}$ is generated by the Hermitian operator $\hat{\epsilon}$ and where the scalar and gauge fields respectively transform in the
fundamental and adjoint representation of the group of unitary operators.
Notice that the action of the operator $\hat{U}$ on $\Phi(X)$ is indeed consistent with the radial dimensional reduction because this unitary operator
preserves the homogeneity degree as $\hat\epsilon$ does.
Notice that as long as higher-derivative transformations are allowed
 then the infinite tower of higher-spin fields should be included for consistency of
 the gauge transformations (\ref{finite}) beyond the lowest order.
The infinitesimal adjoint action \eqref{finite} of the Lie algebra of Hermitian operators on $(A)dS_d$, written in terms of the Weyl symbol $H(X,P)$, leads to the following deformation of \eqref{ambgtransfoo}
\begin{equation}
\delta_\epsilon H(X,P)\,=\,[H_0(X,P)+H(X,P) \, \stackrel{\bigstar}{,} \, \epsilon(X,P)]\,+\,{\cal O}(H^2)\,.
\end{equation}
The ambient generating functions of Killing fields $\overline\epsilon(X,P)$ are Weyl symbols that are product of $X_{[A}P_{B]}$, whose corresponding operators are products of the isometry generators $X_{[A}\partial_{B]}$ of $(A)dS_d$, \textit{i.e.} generators of the Vasiliev off-shell higher-spin algebra. 
When the latter algebra acts on the singleton module of $\mathfrak{o}(d-1,2)$, the three $\mathfrak{sp}(2)$-constraints mentioned at the end of Subsection \ref{symbolcalc} act trivially. The quotient of the Vasiliev off-shell algebra by the corresponding two-sided ideal (spanned by elements
that are sum of elements proportional to a $\mathfrak{sp}(2)$-constraint) is the Vasiliev on-shell higher-spin algebra (see \textit{e.g.} \cite{Vasiliev:2004qz} for more details). The situation is somewhat different for the massive scalar field module spanned by the harmonic homogeneous functions on the ambient space of Subsection \ref{Beltrami}, because this module is not annihilated by the operators corresponding to $X^2$ and $X\cdot P$ (see \textit{e.g.} the section 3 of \cite{Bekaert:2008sa} for some discussion on the algebra of symmetries of the massive scalar field).

It is very tempting to conjecture that the full
action (\ref{actionexpansion}) should be interpreted as arising from the
gauging of the rigid symmetries of the free scalar matter
field, which generalize the $\mathop{\rm {}U}(1)$ and isometries of $(A)dS_d$, so that
the local symmetries (\ref{finite}) generalize the local $\mathop{\rm {}U}(1)$
and diffeomorphisms (see \cite{Calimanesti,Fotopoulos,Bekaert:2008sa,Bekaert:2009ud} and refs therein for more comments
on this point of view). In any case, the unfolded equations (on-shell \cite{Vasiliev:2004qz} and off-shell \cite{V}) precisely arise from the gauging of the same rigid algebra of (on/off shell) symmetries but the scalar field is included in the gauge field multiplet.

To end up with a side remark, we would like to point out the possibility to have a uniform treatment of the gauge fields and parameters where both generating functions have equal homogeneity degree in $X$ and in $P$. This possibility might prove to be useful for further works because this treatment allows to make use of the star commutator induced on ${\cal M}_d$ \cite{Bayen:1977ha} in order to write down the intrinsic form of the gauge transformation \eqref{ambgtransfoo}. Moreover a uniform treatment of fields and parameters is appealing in the metric-like approach since their generating functions can both be interpreted as Weyl symbols of Hermitian (pseudo)differential operators on the spacetime manifold.
Concretely, notice that $\mathcal H(X,P):=X^2H(X,P)$ obeys to 
\begin{equation}
\left(X^A \frac{\partial}{\partial X^A}\,-\,P_A\frac{\partial}{\partial P_A}\,\right)\mathcal H(X,P)\,=\, 0\,,\qquad
\left(X\cdot \frac{\partial}{\partial P}\right)\mathcal H(X,P)\,=\, 0\,,
\end{equation}
as follows from \eqref{hooXP}. The same holds for
\begin{equation}
\mathcal H_0(X,P)\:=\frac12\,X^2\,\left(G^{AB}P_AP_B\mp\,\frac{(mR)^2}{X^2}\right)=\frac12\left(X^2\,P^2\,-\,(X \cdot P)^2\mp(mR)^2\right)
\label{AdSbckgd}
\end{equation}
which corresponds to the Weyl symbol $\frac{\pm R^2}{2}g^{\mu\nu}p_\mu p_\nu\,$.
One can check that
\begin{equation}
[\, \mathcal H_0(X,P)\,\stackrel{\bigstar}{,}\,\epsilon(X,P)\,\,]\,
=\,\left(\,X^2\,+\,\frac14\,{\frac{\partial}{\partial P}} \cdot {\frac{\partial}{\partial P}}\,\right) (P^A\partial_A)\,\epsilon(X,P)
\end{equation}
by making use of the identity
\begin{equation}
2\,[\, X^2\,P^2\,-\,(X \cdot P)^2\,\stackrel{\bigstar}{,}\,\epsilon(X,P)\,\,]\,=\,X^2\star[\,P^2\,\stackrel{\bigstar}{,}\,\epsilon(X,P)\,\,]\,+\,[\,P^2\,\stackrel{\bigstar}{,}\,\epsilon(X,P)\,\,]\star X^2\,.
\end{equation}
Therefore the star commutator between the $(A)dS_d$ background field $g^{\mu\nu}p_\mu p_\nu\,$ and any function $\varepsilon(x,p)$ on the cotangent bundle $T^*(A)dS_n$ above is equal to 
\begin{equation}
\frac12\,\,[\,g^{\mu\nu}p_\mu p_\nu \stackrel{*}{,}\, \varepsilon(x,p)\,] \,=\,
\left(\,1\,\pm\,\frac1{4\,R^2}\,g_{\mu\nu}\,{\frac{\partial}{\partial p_\mu}}{\frac{\partial}{\partial p_\nu}}\right)(p^\mu \nabla_\mu)\,\varepsilon(x,p)\,. 
\end{equation}
Therefore, modulo the field redefinition,
\begin{eqnarray}
h'(x,p)&=&\left(\,1\,\pm\,\frac1{4\,R^2}\,g_{\mu\nu}\,{\frac{\partial}{\partial p_\mu}}{\frac{\partial}{\partial p_\nu}}\right)
\,\,h(x,p)\,,
\end{eqnarray}
the lowest order
of the gauge transformation \eqref{Fronsdalgtransfo} can be expressed directly via the star product on $(A)dS_d$
\begin{eqnarray}
\delta_\varepsilon h'(x,p)\,&=&\,\frac12 \, [\,g^{\mu\nu}p_\mu p_\nu \stackrel{\star}{,}\, \varepsilon(x,p)\,]\,+\,{\cal O}(h')
\end{eqnarray}
in analogy with \eqref{ambgtransfoo}.


\section{Conclusion and outlook}\label{conclusion}

The present paper may be thought as a first step towards a generalization to any constant-curvature spacetime of the results obtained in \cite{Bekaert:2009ud} for a complex scalar field around Minkowski spacetime, such as the generating functions of conserved currents, of interaction vertices, of gauge symmetry deformations and of four-point exchange amplitudes. Besides the exchange amplitudes, all these results have been generalized here to the case of non-vanishing curvature.
Recently, the results of \cite{Bekaert:2009ud} were considerably extended via string-based computations by Sagnotti and Taronna \cite{Taronna:2010qq} and it would be interesting to investigate the possibility of a radial dimensional reduction of their elegant results, looking for the analogue of their generating functions to (anti) de Sitter spacetimes.
We plan to return to these issues in the future.

The generating function of the infinite set of conserved currents for a free complex scalar field on $(A)dS_d$ have been obtained from the flat one \cite{Bekaert:2009ud} through a radial dimensional reduction. For this purpose, an efficient translation rule between ambient partial derivatives and intrinsic (\textit{i.e.} spacetime) covariant derivatives was developed. The form of the current generating function on ambient space is identical to the bilocal function introduced by Fronsdal \cite{Fronsdal:1978vb} in order to provide a manifestly covariant realization of the theorem \cite{Flato:1978qz} asserting that the tensor product of two scalar singleton on the conformal boundary decomposes as an infinite tower of bulk gauge fields. This similarity is by no mean accidental since the Flato-Fronsdal theorem is known to be instrumental in the holographic correspondence between free conformal field theories on the boundary and higher-spin gauge field theories in the bulk 
but it might deserve to be investigated further in the ambient formulation.

Through the Noether method, the current generating function allows to write a generating function of cubic minimal couplings and to determine the corresponding gauge symmetry deformations.
Our results confirm some previous expectations on the non-Abelian deformation of the metric-like gauge symmetry as being the group of unitary operators on the spacetime manifold, thereby generalizing the group of diffeomorphisms. 
It was extremely convenient to remove trace constraints on the gauge parameters when reflecting on the non-Abelian symmetries in the metric-like formulation of higher-spin gauge fields (see \textit{e.g.} \cite{Bekaert:2008sa} for an extended discussion of this point). 
As far as the non-Abelian frame-like formulation is concerned, the analogue of
Vasiliev's unfolded equations in the unconstrained case are also of interest for studying the off-shell gauge symmetry structure \cite{V}. 
Moreover, a slight refinement of the on-shell unfolded equations has been proposed in \cite{SSS} following the spirit of the unconstrained approach.
The recent frame-like formalism with weaker trace constraints \cite{Sorokin:2008tf} might also shed some light in these directions.

Notice that, at the order where we worked (at most quadratic dependence in the gauge fields), it is perfectly consistent to make use of traceful currents in the ``minimal'' coupling between gauge fields and currents. However, the quadratic action for the gauge fields will determine the genuine physical interactions between the matter and gauge fields. Indeed, the gauge fields may also couple to other fields, dynamical or not (\textit{e.g.} auxilliary and pure gauge fields), and these couplings will affect the on-shell structure of the interactions.
For instance, if the quadratic gauge field action is the Fronsdal action \cite{Fronsdal:1978vb} then the double-trace of the current is automatically extracted out off-shell and the single-trace further decouples on-shell. It is known since the seminal works of Francia and Sagnotti that the trace constraints may consistently be removed off-shell from the metric-like quadratic action in several ways for irreducible gauge fields (see \textit{e.g.} \cite{FS} for some reviews and \cite{Francia:2008hd,BGK} for some recent developments). Nevertheless, the trace of the current still decouples on-shell as it should \cite{Francia:2008hd}.
For the so-called ``triplet'' arising from the open string leading Regge trajectory \cite{FS,Francia:2010qp} (see also \cite{Fotopoulos:2007yq,Fotopoulos:2010nj}), the situation is more subtle: although traceful
conserved currents can indeed source the symmetric tensor field, only the traceless component of the currents
studied here leads to genuine minimal interactions.\footnote{We are grateful to the referee for calling this fact to our attention.} The $k$th trace of the current of rank $r$ is a current of rank $r-2k$  (lower than $r$) and contains $r$ derivatives. However, any non-trivial rank-$s$ conserved current built from a scalar field is known to contain up to $s$ derivatives.
Therefore, any trace component of the current is equal on-shell either to zero or to an ``improvement'', \textit{i.e.} a trivially conserved (or, equivalently, co-exact) current. Such on-shell trivial currents give rise to non-minimal interactions, quadratic in the scalar fields and linear in the gauge-invariant higher-spin fieldstrengths.

Finally, the toy model \cite{Bekaert:2009ud} has been used to calculate tree level exchange amplitudes
for the elastic scattering of two scalar particles mediated by an infinite tower of tensor gauge fields.
The $AdS_{d}$ counterparts of Feynman diagrams with four external scalar particles should be Witten diagrams associated with the four-point correlation function of a singlet (``single trace'') scalar operator, bilinear in some large component massless scalar field living on the conformal boundary, as in \cite{Sezgin:2002rt,Klebanov:2002ja}. 
The exact summation of the corresponding exchange amplitudes
for an infinite tower of intermediate tensor gauge fields is possible in flat spacetime \cite{Bekaert:2009ud} and one might hope to reproduce the analogue of this result in $AdS_{d}$ since all ingredients are now available in the unconstrained formalism for irreducible gauge fields: the bulk-to-bulk propagators of symmetric tensor fields can be extracted from \cite{Francia:2008hd} and the relevant cubic vertices have been presented here.\footnote{The analogue of these cubic vertices were obtained in the \textit{constrained} formalism by Fotopoulos, Irges, Petkou and Tsulaia \cite{Fotopoulos}. However, we believe that, as suggested by the case of flat spacetime, the unconstrained formalism could be technically more handy for Feynmann/Witten diagram computations.} Moreover, the $CFT_{d-1}$ dual results are known in closed form, even for the interacting $O(N)$ model in the large $N$ limit \cite{Leonhardt:2002sn}.
Computing explicitly the $AdS_d$ exchange Witten diagram could therefore provide a first quantitative test of the $AdS_4/CFT_3$ conjecture of Klebanov and Polyakov \cite{Klebanov:2002ja} at quartic level, \textit{i.e.} for four-point correlation functions. Indeed, while impressive quantitative checks of the correspondence have been performed at the interacting level \cite{Petkou:2003zz,Sezgin:2003pt,Giombi:2009wh}, to our knowledge all of them were restricted yet to three-point correlation functions where symmetries are known to highly constrain the set of possibilities.

\acknowledgments

We thank N. Boulanger, E. Joung, J. Mourad, A. Sagnotti, P. Sundell, M. Taronna and especially S. Nicolis for useful exchanges.


\appendix

\section*{Technical appendix}

Let us consider a homogeneous symmetric tensor of rank $r$ such that $(X^B\partial_B-k)\,T_{A_1\ldots A_r}(X) \,=\,0$ and $X^{A_1}\,T_{A_1\ldots A_r}(X) =0$.
These last two properties together with the definitions of the projector \eqref{projector} and the equivalence relation \eqref{equrel} imply that
\begin{eqnarray}
&&\partial_A \Big(\,\mathcal P^{D_1}_{B_1}\ldots \mathcal P^{D_n}_{B_n}\mathcal P^{E_1}_{C_1}\ldots\mathcal P^{E_r}_{C_r}\partial_{D_1}\ldots \partial_{D_n} T_{E_1 \ldots  E_r}\Big)\sim
\nonumber\\
&&\qquad\sim\,\partial_A\partial_{B_1}\ldots\partial_{B_n}T_{C_1\ldots C_r}\,-\,\frac1{X^2}\,n\,(X^D\partial_D)\,\eta_{A(B_1}\,\partial_{B_2}\ldots \partial_{B_n)} T_{C_1 \ldots C_r}\,-
\nonumber\\
&&\qquad\qquad-\,\frac1{X^2}\,r\,X^E\,\partial_{B_1}\ldots \partial_{B_n} T_{E(C_2 \ldots C_r}\eta_{C_1)A}
\nonumber\\
&&\qquad=\,\partial_A\partial_{B_1}\ldots\partial_{B_n}T_{C_1\ldots C_r}\,-\,\frac1{X^2}\,n\,\big(k-(n-1)\big)\,\eta_{A(B_1}\,\partial_{B_2}\ldots \partial_{B_n)} T_{C_1 \ldots C_r}\,+
\nonumber\\
&&\qquad\qquad+\,\frac1{X^2}\,r\,n\,\partial_{(B_2}\ldots \partial_{B_n}T_{B_1)(C_2 \ldots C_r}\eta_{C_1)A}
\nonumber
\end{eqnarray}
Contracting all indices with an auxiliary vector $P$ and making use of the notations  \eqref{explcovder} and \eqref{notat}, one gets that
\begin{eqnarray}
\label{rhslhs}
 (P\cdot\mathcal D)\,(P\cdot\partial)^nT\,=\, (P\cdot\partial)^{n+1}\,T\,-\,n\,\dfrac{P^2}{X^2}\,\big(k-r-(n-1)\big)\,(P\cdot\partial)^{n-1}\,T
\end{eqnarray}
The left-hand-side of \eqref{rhslhs} can be expressed by
\begin{eqnarray}
\nonumber
(P\cdot\mathcal D)\,(P\cdot\partial)^n\,T&=& (P\cdot\mathcal D)\,\sum \limits_{m=0}^{[n/2]}c_n^m\,\left(\dfrac{P^2}{X^2}\right)^r \, (P\cdot\mathcal D)^{n-2r}\,T\\
\nonumber
&=& c_n^0\,(P\cdot\mathcal D)^{n+1}\,+\,\sum \limits_{m=1}^{[n/2]}c_n^m\,\left(\dfrac{P^2}{X^2}\right)^m \, (P\cdot\mathcal D)^{n-2m+1}\,T\,.
\end{eqnarray}
where \eqref{def} has been inserted in order to compute $(P\cdot\partial)^n$.
The right-hand-side of \eqref{rhslhs} can also be reexpressed as follows
\begin{eqnarray}
\nonumber
&&(P\cdot\partial)^{n+1}\,T\,-\,n\,\dfrac{P^2}{X^2}\,(k-r-(n-1))\,(P\cdot\partial)^{n-1}\,T \\
\nonumber
&=&c_n^0\,(P\cdot\mathcal D)^{n+1}\,T\,+\,\sum \limits_{m=1}^{[(n+1)/2]}\left(\dfrac{P^2}{X^2}\right)^m \, (P\cdot\mathcal D)^{n+1-2m}\,[c_{n+1}^m\,-n(k-r-n+1)\,c_{n-1}^{m-1}] \,T\,.\end{eqnarray}
by making use twice of \eqref{def} in order to calculate $(P\cdot\partial)^{n+1}$ and $(P\cdot\partial)^{n-1}$.
These equations imply that the coefficients $c_n^m$ are given by the recurrence formula:
\begin{eqnarray}
c_{n+1}^m\,=\,c_n^m \,+\,n\,(k-r-n+1) \,c_{n-1}^{m-1}
\label{doublerec}
\end{eqnarray}
and for $n$ odd, there is an additional relation:
\begin{equation} c_{n+1}^{(n+1)/2}\,=\,n(k-r-n+1)\,c_{n-1}^{(n-1)/2}\,.\end{equation}

\noindent If one considers the $c_n^m$ as the coefficients of a power (\textit{a priori} formal) series \begin{equation}f(x;y) \,=\,\sum\limits_{n=0}^{\infty} \,\sum\limits_{m=0}^{\infty}\, \frac{1}{n!} \, c_n^m \, x^n\,y^m\,,\end{equation} one can rewrite the recursion formula \eqref{doublerec} as an ordinary differential equation (parametrized by the ``constant'' $y$) for the unkown function $f(x,y)$ depending on the single variable $x$ 
\begin{equation}(1\,+\,x^2\,y)\,\frac{d}{dx}\,f(x,y)\,-\,\big(1\,+\,(k-r)\,xy\big)\,f(x,y) \,=\,0 \end{equation}
with the initial condition $f(0,y)\,=\,1$.
The solution of this Cauchy problem is : \begin{equation}f(x,y)\,=\,(1\,+\,y\,x^2)^{\frac{k-r}2}\,\exp\left( \frac{1}{\sqrt{y}}\,\arctan\,(\sqrt{y}\,x)\right).\end{equation}
The generating function $c(x,y;k)$ is equal to $f(x,y/x^2)$.

\TABLE[thp]{
\begin{tabular}{|c|c|c|}
\hline
Object  & Ambient space $\mathbb{R}_0^D$ & Constant-curvature spacetime $\mathcal{M} _d$  \\
\hline
Coordinates & $X^A$ &  $x^\mu$  \\
\hline
Scalar & $\Phi(X)$ & $\phi(x)$ \\
\hline
Conjugate & $\Phi^\dagger(X)$ & $\phi^*(x)$ \\
\hline
Vector &  $T_A(X)$ &$t_\mu(x)$ \\
\hline
Tensor & $T_{A_1\ldots A_r}(X)$ &  $t_{\mu_1\ldots \mu_r}(x)$  \\
\hline
Metric & $G_{AB}\,\sim\,\eta_{AB}$ & $g_{\mu\nu}$  \\
\hline
Covariant derivative &  $\mathcal{D}_A$ &   $\nabla_\mu$  \\
\hline
Spacetime Laplacian& $\mathcal{D}^2\,=\,G^{AB}\,\mathcal D_A \,\mathcal D_B$ & $\nabla^2\,=\,g^{\mu\nu}\,\nabla_\mu\,\nabla_\nu$  \\
\hline
Ambient Laplacian & $\partial^2\,=\,\eta^{AB}\,\partial_A\,\partial_B$ & $\nabla^2\,\pm\,\dfrac{1}{R^2}\,k\left(k\,+\,d\,-\,1\right)$ \\
\hline
\end{tabular}
\caption{Dictionary Ambient space/Constant-curvature spacetime} 
\label{dico2} 
} 

\end{document}